\documentclass[12pt]{article}

\usepackage[dvipsnames, svgnames, x11names]{xcolor}
\usepackage{scicite,graphicx}
\usepackage{times}
\usepackage{sci_macros}
\usepackage[normalem]{ulem}
\usepackage{soul}
\usepackage{url}
\usepackage[colorlinks]{hyperref}
\usepackage{subfig}
\usepackage[font=footnotesize]{caption}

\usepackage{floatrow}
\DeclareFloatFont{tiny}{\scriptsize}
\makeatletter

\newenvironment{sciabstract}{%
\begin{quote} }
{\end{quote}}

\newcounter{firstbib}

\begin{document}

\title{PeV $\gamma$-rays from the Crab}
\date{}
\maketitle
\centerline{\large The LHAASO Collaboration$^{\star\dag}$}
\vspace{10pt}
\noindent $^\star$Collaboration authors and affiliations are listed in the Supplementary Material\\
$^\dag$Correspondence to:  
caozh@ihep.ac.cn; chensz@ihep.ac.cn; linsj6@mail.sysu.edu.cn;\\
zhangss@ihep.ac.cn; zham@ihep.ac.cn; licong@ihep.ac.cn; wangly@ihep.ac.cn;\\
yinlq@ihep.ac.cn; felix.aharonian@mpi-hd.mpg.de; ryliu@nju.edu.cn

\baselineskip24pt
\maketitle
\clearpage

\begin{sciabstract}
The Crab pulsar and the surrounding nebula powered by the pulsar’s 
rotational energy through the formation and termination of a 
relativistic  electron-positron wind is a bright source of gamma-rays 
carrying crucial information about this complex conglomerate. We report 
the detection of $\gamma$-rays with a spectrum showing gradual
steepening over three energy decades, from $5\times10^{-4}$ to 1.1 
petaelectronvolt (PeV).  The ultra-high-energy photons exhibit the 
presence of a PeV electron accelerator (a pevatron) with an 
acceleration rate exceeding 15 \% of the absolute theoretical limit.
Assuming that unpulsed $\gamma$-rays are produced at the termination of 
the pulsar’s wind, we constrain the pevatron’s size, between 0.025 and 
0.1 pc, and the magnetic field, $\approx 110 \mu$G. The production rate 
of PeV electrons,  $2.5 \times 10^{36}  \rm erg  \ s^{-1}$, constitutes 
0.5\% of the pulsar’s spin-down luminosity, although we do not exclude a 
non-negligible contribution of PeV protons to the production of the 
highest energy $\gamma$-rays.
\end{sciabstract}
\noindent

\newpage

\section*{Main text:}

{
The Crab Nebula is a remnant of the 
explosion of a massive star, formed on Aug 24, 1054 A.D., as  recorded in Chinese chronicles as a  `guest star' \cite{gueststar}. It is the brightest pulsar wind nebula, an extended nonthermal structure  powered by 
the ultrarelativistic electron-positron wind from the central neutron star (the Crab Pulsar).  
This makes the Crab Nebula one of the brightest $\gamma$-ray sources in the sky, which has been observed for decades 
at TeV energies\cite{Whipple-on-crab}.
At the transition from GeV to TeV energies, the spectral energy distribution (SED; $E^2 {\rm d}N/{\rm d}E$, where  ${\rm d}N/{\rm d}E$ is the differential flux of radiation 
at the photon energy $E$) achieves 
a maximum around 100 GeV \cite{Arakawa20,Magic-crab}.
Previous observations reported a detection of $\approx 80$~TeV\cite{HEGRA-crab}, 
and 
measured the spectrum up to  
300~TeV 
\cite{magiccrab2020,HAWC_crab,Tibet_crab}.
The angular size of the $\gamma$-ray source at TeV energies has been 
reported $\approx 50$~arcsec\cite{HESSCrabsize}. 
The Crab Nebula was among the first sources detected at energies well beyond 100~TeV using Large High Altitude Air Shower Observatory (LHAASO) \cite{12sourcespaper}. 
}

{
LHAASO is a dual-purpose complex of particle detectors 
designed for the study of cosmic rays (CRs) and $\gamma$-rays in the sub-TeV to 1000 PeV energy range\cite{Cao10,He18}.  When a 
very high energy extraterrestrial particle
enters Earth's atmosphere, it initiates a cascade  
consisting of secondary hadrons, leptons, and photons, known as an air shower. The LHAASO detectors record different components of air showers which are used to reconstruct the type, energy and arrival direction of the primary particles. 

LHAASO consists of three arrays \cite{SM}. 
The largest 
is the square kilometer array (KM2A)  composing surface 
counters and subsurface muon detectors. KM2A is designed to detect  cosmic rays (CR)  from  10 TeV to 100 PeV and identify $\gamma$-ray photons among the CRs using  
the muon detector array. The muon content in an air shower event measured by later can be used to  effectively reject hadronic showers initiated by CRs. The surface array of KM2A is used to measure the energy and arrival direction of primary particles. Thus, KM2A serves as an Ultra High Energy (UHE, $E_\gamma > 0.1$ PeV) $\gamma$-ray telescope\cite{KM2A-on-Crab} with energy resolution of $\leq 20 \%$ angular resolution
and 0.25$^\circ$, respectively. For $\gamma$-rays above 50 TeV, the
 sensitivity of KM2A reaches the flux level of $10^{-14} \rm erg~ cm^{-2} s^{-1}$ for a single source like the Crab Nebula per year\cite{He18}.  
 }
 
The Water Cherenkov Detector Array (WCDA) 
consists of  three ponds with total area of $0.08 \ \rm km^2$, 
sensitive to $\gamma$-rays down to 0.1~TeV 
with an angular resolution 
$\delta \psi \approx 0.2^\circ$\cite{WCDA-on-Crab}.
WCDA is designed to perform 
deep surveys 
{of very high energy  $\gamma$-ray sources, including the Crab for both pulsed and unpulsed signals. The sensitivity reaches the flux level of $10^{-12} \rm erg~ cm^{-2} s^{-1}$ for a single source like the Crab Nebula per year at energies above 2 TeV\cite{He18} . 
At energies above 0.1 PeV, WCDA also serves as a large muon detector to enhance the capability of separation between electromagnetic and hadronic showers detected by KM2A.  
}

KM2A and WCDA are complemented by the  Wide-Field-of-view Cherenkov Telescope Array  (WFCTA)\cite{WFCTA-NIM} consisting of 18 telescopes designed for detection of the Cherenkov radiation
emitted by air showers induced  by CRs 
with energy ranging from 0.1 to 1000 PeV. 
{Cherenkov light in showers initiated by $\gamma$-rays at energies above 0.1 PeV is recorded by the currently operating fourteen WFCTA telescopes} which cover a $32^\circ\times 112^\circ$  patch of the northern sky through which the Crab passed every day.

\noindent{\bf  LHAASO Observations of the Crab}

\begin{figure}
  \centering\includegraphics[width=1.0\textwidth]{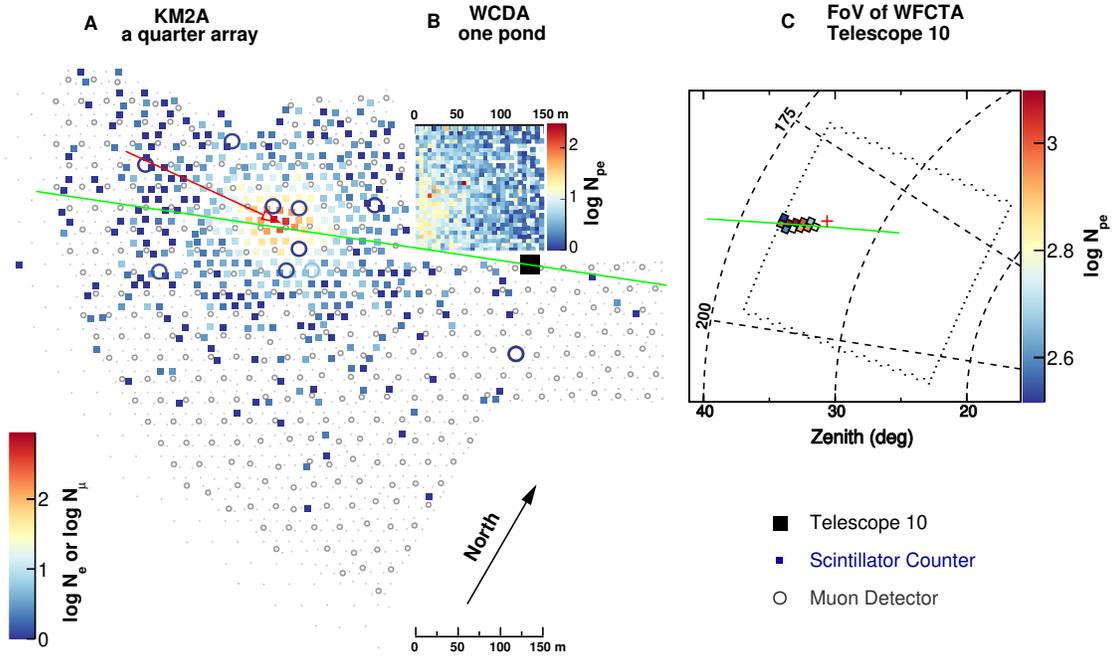} 
  \caption{{\bf The 0.88~PeV $\gamma$-ray event from the Crab recorded by the LHAASO detectors. }  In panel A,
  squares 
  indicate the scintillator counters of KM2A, colored according to the logarithm of number of 
  detected particles  
  $N_e$ (color bar). 
  The open circles indicate the 11 Muon Detectors of KM2A triggered by the shower.  
  The position of the core is indicated by the red arrow, which is orientated in the arrival direction of the primary photon.  The panel B shows the map of WCDA detector units. The logarithm of the number of photoelectrons recorded in each unit is indicated by the color. The scale is represented by the bar. On the southern side of WCDA, the  Telescope-10 of WFCTA which also detected the event. 
  The panel C shows the telescope FoV  outlined by the dotted lines, while the dashed arcs indicate zenith angles of   20$^\circ$, 30$^\circ$, 40$^\circ$, from right to left and dashed lines indicate azimuth angles of 175$^\circ$, 200$^\circ$ counterclockwise. The shower image, composed of 11 pixels, started about 34$^\circ$ in zenith and stretched to the edge of the FoV at 38$^\circ$. 
  The colour scale shows the logarithm of the number of photoelectrons in each pixel.  The main axis of the image in the FoV of the telescope indicates the shower-telescope-plane which is consistent with the event direction (indicated by the red cross in panel C) reconstructed using KM2A. The green line in panel A 
  is the intersection of the shower plane and the ground is consistent with the shower core, as well.  
  }
  \label{fig:maxE-event}
\end{figure}

On 2020 January 11 at 17:59:18 Coordinated Universal Time (UTC),  a giant air shower was recorded by all three LHAASO detectors. The shower arrived from the direction of Crab and landed in the western part of KM2A (see Fig.\ref{fig:maxE-event}).  
WCDA was triggered, despite being 
150~m away from the shower core.  The event occurred after local midnight when eight WFCTA telescopes were operational. The Cherenkov radiation of the air shower appeared in the Field of View (FoV) of Telescope No.10 and triggered it
(also see Fig.\ref{fig:maxE-event}).

\begin{figure}
 \centering\includegraphics[width=0.8\linewidth]{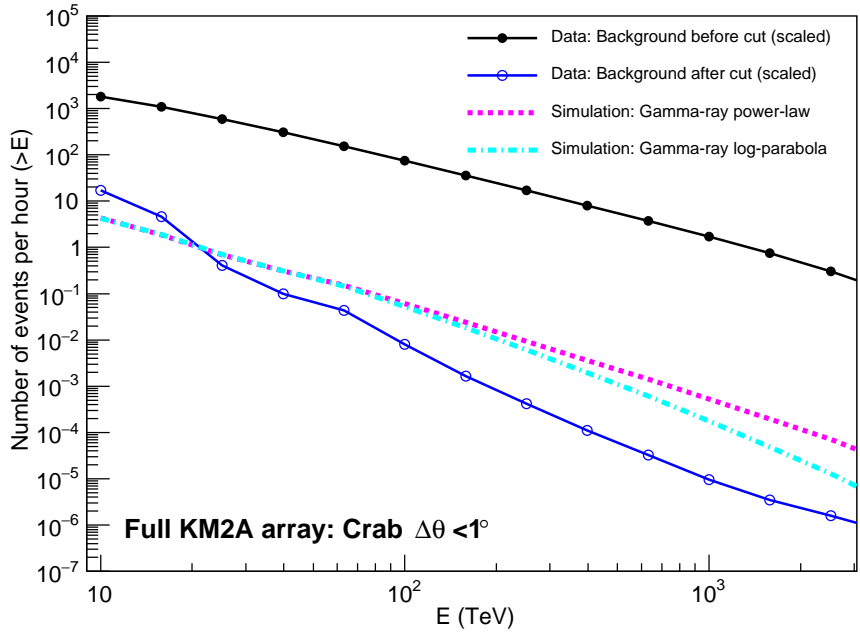}
  \caption{ {
  {\bf The rates of detection of $\gamma$-rays from the Crab and 
  the cosmic ray background events above the shower energy $E$ by the 
  $1 \ \rm km^2$ array in a cone of 1$^\circ$ centered at the Crab direction.} The rates correspond to the number of events per hour of observation of the source within the FoV of KM2A. 
  $E$ is the reconstructed shower energy.
  The cyan dash-dotted and pink dashed lines represent the integrated detection rates   of $\gamma$-rays from the Crab,  based on   log-parabola and power-law models fitted to the measured fluxes (see Fig.\ref{fig:Crab-SED}), respectively. 
The black filled circles show the rate of cosmic ray events accumulated by KM2A. The integrated rate can be approximately described  as  power law with index of -1.6. 
  The blue open circles  represent  the integrated rate of cosmic ray events after the `muon cut' filter.  
  }
  }
  \label{fig:BG-event-cut}
\end{figure}

\begin{figure}
  \centering\includegraphics[width=0.8\linewidth]{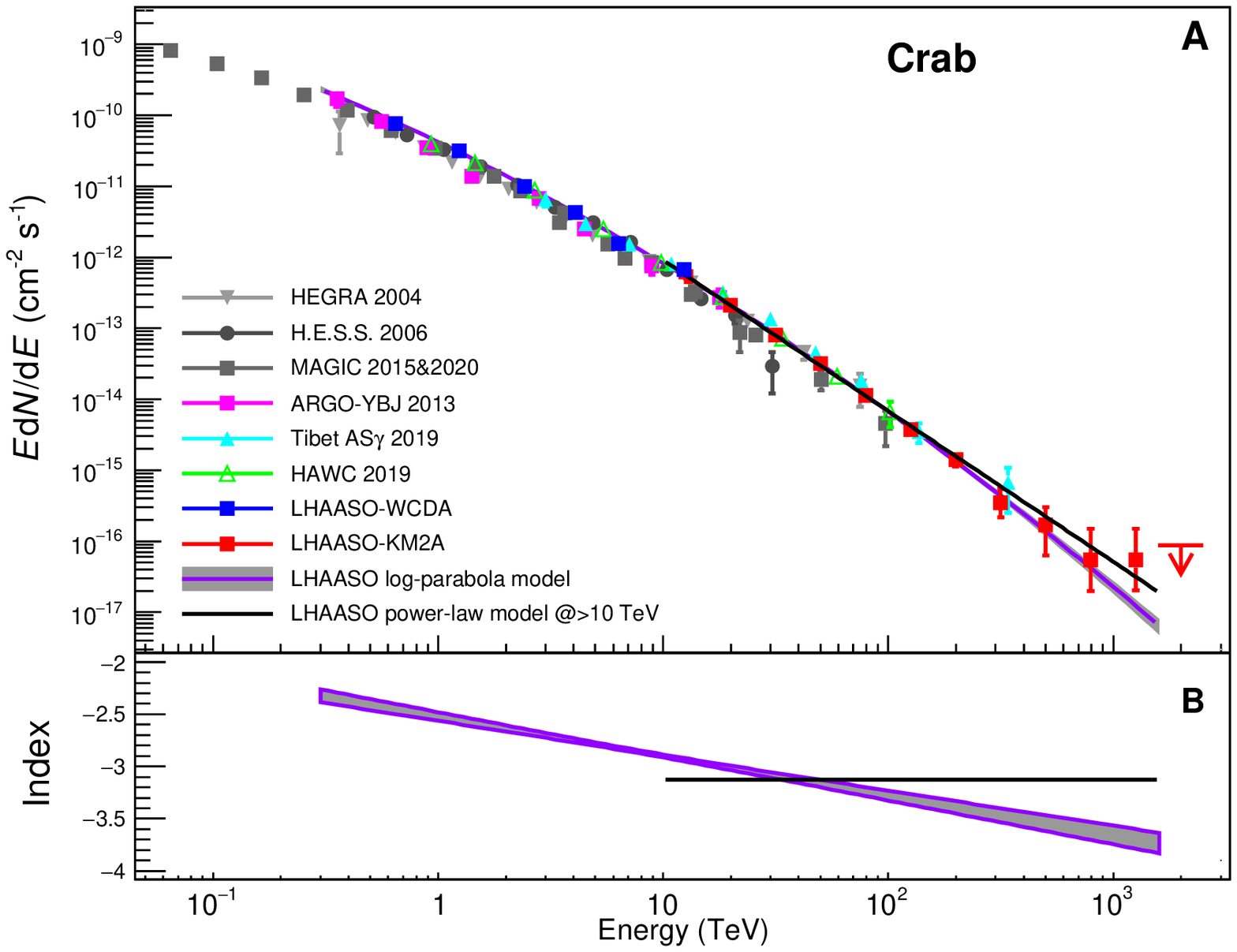}
    \caption{
 { {\bf $\gamma$-ray flux of the Crab measured by LHAASO and spectral fitting.} Panel A shows TeV to PeV $\gamma$-ray fluxes of the Crab plotted as
   $ E {\rm d}N/{\rm d}E$. 
  The red squares and blue squares are the spectral points measured using KM2A and WCDA, respectively. The spectral points above 100 TeV were obtained in the  signal-dominated  regime, with 89 detected $\gamma$-rays and 2 events expected from   CR induced (hadronic) air showers   after the  muon  cuts. No events were detected  in the 1.6 to 2.5 PeV bin where an arrow indicates the flux upper limit at 90\% confidence level. 
  The purple line shows the fitting  using a log-parabola (LP) model in the  0.3~TeV to 1.6~PeV interval ($\chi^2/dof$: 9.3/14). For comparison, the black line shows   the fitting using a simpler power-law (PL) model in the  10~TeV to 1.6~PeV interval ($\chi^2/dof$: 5.4/9). 
  Also plotted are previous observations of the Crab by other facilities: HEGRA\cite{HEGRA-crab}, H.E.S.S.\cite{HESS-crab}, MAGIC\cite{Magic-crab,magiccrab2020}, ARGO-YBJ\cite{ARGO_crab}, HAWC\cite{HAWC_crab}, Tibet AS$\gamma$\cite{Tibet_crab}. Panel B shows the energy-dependent local power-law index $\Gamma$  derived by the  log-parabola model fitting, as indicated by the purple band. For comparison, the black line shows  the photon index $3.12 \pm 0.03$ derived from the simpler power-law model fitting.  Error bars represent one standard deviation.
  }
  }
  \label{fig:Crab-SED}
\end{figure}

{
We identify the event as a $\gamma$-ray induced shower based on 4996 particles (electrons, photons, muons, hadrons)  recorded by 395 surface detectors and 15 muons recorded by 11 under-surface detectors of KM2A. The chance probability of this event to be misidentified 
is estimated as 0.1\%.
Two independent estimates of the shower energy were
derived from the KM2A and WFCTA data are 
$0.88\pm 0.11$~PeV and $0.92^{+0.28}_{-0.20}$~PeV, respectively. This value has been reported as the maximum energy of $\gamma$-rays detected from the Crab direction by LHAASO elsewhere\cite{12sourcespaper}.

Approximately one year later, 2021 January 4 at 16:45:06, another shower at even higher energy, i.e. 1.12$\pm0.09$~PeV, was registered by KM2A at zenith angle  $12.9^\circ$ which is more vertically arrived and better measured than the previous one by KM2A. Unfortunately, the primary photon arrived one hour before the Crab entered the  FoV of the WFCTA telescopes. While the number of detected secondary particles (5094) exceeded those in the previous event, the number of muons (14) was fewer, we also identified this event as a $\gamma$-ray induced shower with a 0.03\% chance probability of mis-identification.

KM2A operated for 314 days at half of its design capacity and a further 87 days at three quarters of its capacity. In that time, a total of 89 UHE $\gamma$-rays with energy exceeding 0.1~PeV were detected from the Crab. }
Fig.\ref{fig:BG-event-cut} shows the 
integrated $\gamma$-ray detection rate, re-normalized for the nominal $1 \ \rm km^2$ array and for one hour exposure time, assuming the Crab is within the FoV of KM2A (approximately 7.4 hours per day with zenith angle less than 50$^\circ$).  Above 0.1 PeV, we find about 0.05 events per hour, equivalent to  
 135 events per year.
Fig.\ref{fig:BG-event-cut} also shows the detection rate of the CR induced showers within a  cone of $1^\circ$   centred on the Crab,  both before and after applying the CR shower rejection based on the muon content in the showers, namely the `muon cut' filter requiring that the number of muons detected by KM2A in the shower must be less than 1/230 of the number of particles detected by the KM2A surface counters. This cuts the cosmic ray background  by the factors of 1,000 and 500,000
at 50 TeV and  1 PeV, respectively. At energies above 0.1 PeV, the detection rate of $\gamma$-rays from the Crab exceeds the CR induced background by an order of magnitude. {
Because the Point Spread Function (PSF) of KM2A is $\delta \psi \approx 0.25^\circ$, 
the CR background could be lower by an additional  factor of $(1^\circ/\delta \psi)^2 \approx 16$.

\noindent{\bf The Spectral Energy Distribution Measurement}

The $\gamma$-ray fluxes in the energy range from 0.5 TeV to 13\, TeV are measured using the firstly built pond of WCDA. From 2019 September to 2020 October, the total exposure was 343.5 transits of the Crab. The measurement using KM2A in the observation period reported above covers the higher energy range from 10\,TeV to 1.6\,PeV. The SED of the Crab are shown in Fig.\ref{fig:Crab-SED}.  The two independent measurements of $\gamma$-ray flux using two detector components of LHAASO fit a simple SED functional form ${\rm d}N/{\rm d}E =(8.2\pm0.2)\times 10^{-14} (E/10 \ \rm TeV)^{-\Gamma}$  cm$^{-2}$s$^{-1}$TeV$^{-1}$, where $N$ is the number of $\gamma$-rays, $E$ is the $\gamma$-ray energy and $\Gamma$ is the energy dependent spectral index. The two measurements are smoothly connected in the small overlapping region around 12.5 TeV. In this energy bin, the discrepancy between flux measured by KM2A and WCDA is 1.3$\sigma$. 
The overall $\chi^2/dof$ is 9.3/14, where $dof$ refers to degrees of freedom. 
No systematic deviation between the two segments of the SED is observed. 
Function $\Gamma = (2.90\pm 0.01) + (0.19\pm0.02)\log_{10} (E/10 \ \rm TeV)$ 
implies a gradual steepening of the spectrum 
characterised by the local index $\Gamma$, from $\approx 2.5$ at 1\,TeV to 3.7 at 1\,PeV.

Fig.\ref{fig:Crab-SED} also shows previous
measurements using atmospheric Cherenkov telescopes 
\cite{HEGRA-crab,HESS-crab,magiccrab2020,VERITAS_CRAB}
and the 
air shower arrays \cite{ARGO_crab,HAWC_crab,Tibet_crab}. 
These are consistent with the WCDA and KM2A data
from sub-TeV to multi-TeV energies, including the flux at the highest energy $E_\gamma \approx 300$~TeV previously reported\cite{Tibet_crab}. 
}

{
Photons from the Crab Nebula have been detected over 22 decades { of energy}, from MHz radio to  UHE $\gamma$-rays, and consists of several pulsed and unpulsed components.  $\gamma$-rays can be produced in three physically distinct sites - in the pulsar's magnetosphere,   ultrarelativistic electron-positron (hereafter, simply electron) wind  and nebula. UHE $\gamma$-rays are absorbed in the strong magnetic field of the pulsar, thus a pulsed component has been expected only from the magnetosphere in the form of MeV-to-GeV $\gamma$-rays. However, the surprise detection of pulsed TeV $\gamma$-rays from the Crab \cite{VHECrab_pulsed} initiated a new concept allowing the location of the pulsed $\gamma$-ray production in the wind \cite{Aharonian12,Mochol,Yeung}.  Although the extension of this component to the  UHE domain seems rather problematic,  we plan to search for a pulsed component in the LHAASO data after adequate photon statistics is accumulated. Below we assume that the entire flux detected by LHAASO consists of an unpulsed component and is produced in the nebula.}

The broad-band nonthermal emission of the Crab Nebula is dominated by two mechanisms -  synchrotron radiation and inverse Compton (IC) scattering of relativistic electrons interacting with the ambient magnetic and radiation fields \cite{DeJagerHarding1992,AA96}. 
In the standard paradigm, the acceleration of electrons is initiated by the termination of the  wind by a standing reverse shock at a distance $R \approx 0.1$\,parsec (pc)  from the pulsar \cite{ReesGunn,KenCor}. 
Although the details of the acceleration remain unknown,  the  detection of  PeV photons allows us to estimate the accelerator size $l$, the magnetic field strength $B$, and the minimum acceleration rate $\eta$.

In the Crab Nebula, several radiation fields supply target photons for the IC scattering of electrons. However, at energies above 100 TeV, the 2.7~K Cosmic Microwave Background radiation (CMBR) dominates the $\gamma$-ray production\cite{AA96,Meyer}. 
Because the CMBR is well quantified, the $\gamma$-ray data provides direct information about the parent electrons.  $\gamma$-ray production above 100~TeV proceeds  in the Klein-Nishina regime, where the energies of the upscattered photon $E_\gamma$ and the  
parent electron $E_{\rm e}$ are linked through the simple
relation  $E_\gamma=0.37 (E_{\rm e}/1 \, \rm PeV)^{1.3} \, PeV$, which over two energy decades, from 30 TeV to 3 PeV, 
provides accuracy better than 10\%  (see Fig.\ref{fig:syn-ic}), or  equivalently  
\begin{equation}
E_{\rm e} \simeq 2.15 (E_\gamma/1 \ \rm PeV)^{0.77} \ \rm PeV \ .
\label{Ee}
\end{equation}
Thus, for the 1.1~PeV photon,  the energy of the parent electron is 2.3~PeV.  Correspondingly, the mean energies of the synchrotron 
($\varepsilon_{\rm syn}$) 
and IC ($E_\gamma$)  photons 
produced by the same electron 
in the ambient magnetic field $B$, 
are related by 
\begin{equation}
\varepsilon_{\rm syn}=9.3 (E_\gamma/1 \, {\rm PeV})^{1.5} (B/100 \mu \rm G) \, \rm MeV \ .
\label{EsyEic}
\end{equation}

\begin{figure}
  \centering\includegraphics[width=1\linewidth]{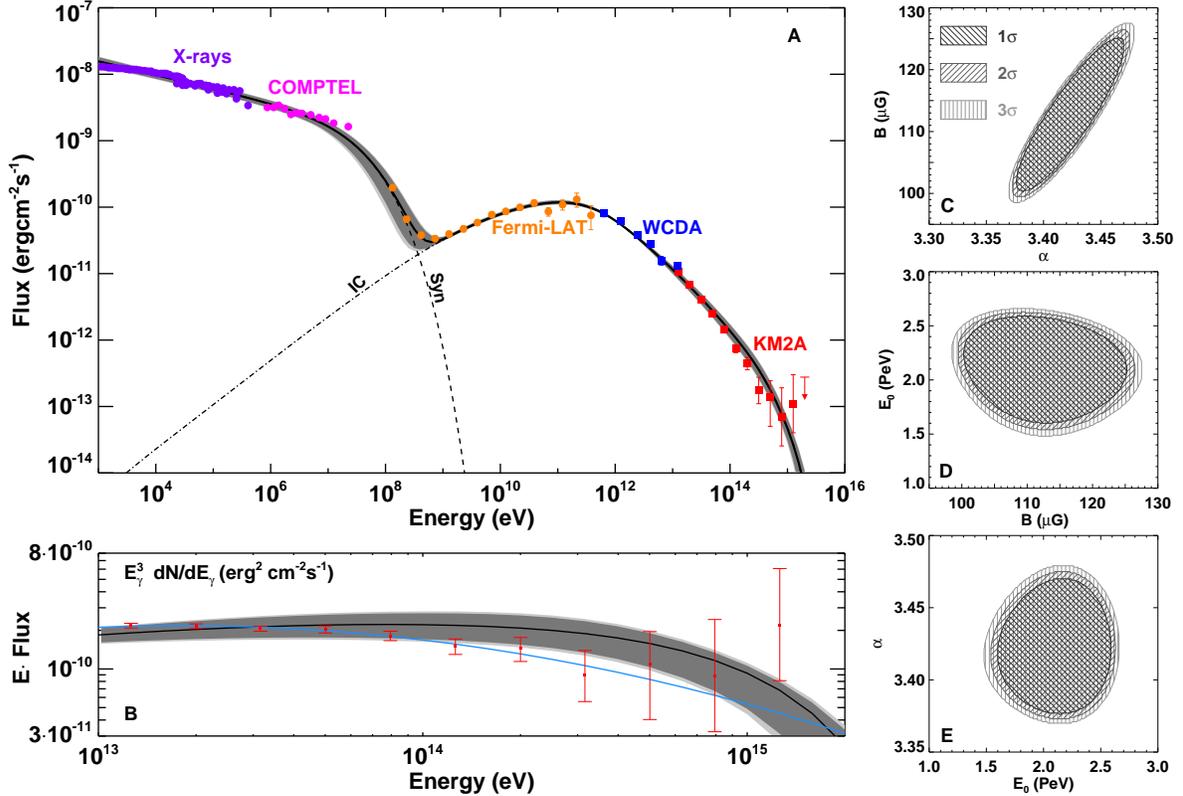}
  \caption{{\bf The Spectral Energy Distribution of the Crab Nebula. } Panel A: The black curves represent the fluxes of the Synchrotron and IC components of radiation of an electron population calculated within the one-zone model. The electron spectrum above 1 TeV is assumed to be 
  a power-law function terminated by an super-exponential cutoff, 
  $E^{-\alpha}\exp[-(E/E_0)^2]$. The model's best fit parameters are: $\alpha=3.42\pm 0.05$, $E_0=2.15_{-0.65}^{+0.55}\,$PeV, $B=112_{-13}^{+15}\,\mu$G. The total energy in electrons above 1\,TeV is $W_e=7.7\times 10^{47}\,$erg. A break in the electron spectrum at $E_b=0.76\,$TeV is assumed to provide
  a consistency with the GeV $\gamma$-ray and low-frequency synchrotron  data (see discussions in SM and Fig.~\ref{fig:Crab-RtoG}). The dark-grey and light-grey shaded regions show the 1$\sigma$ and 3$\sigma$ error bands, respectively. The purple and the magenta circles show the X-ray and the MeV emission of the Crab Nebula\cite{Kuiper01}. The orange circles represent the Crab observations by Fermi-Large Area Telescope (LAT) in the non-flare state\cite{Arakawa20}. The blue and red squares represent WCDA and KM2A measurements reported in this work.  
  Panel B zooms the fluxes above 10\,TeV in the presentation of $E^3{\rm d}N/{\rm d}E$. The blue curve presents the log-parabola spectral fitting shown in in Fig.\ref{fig:Crab-SED}. 
  Panel C, D and E show the 2-dimensional projected parameter spaces of the free parameters $\alpha$, $B$ and $E_0$.
  }
  \label{SED-onezone}
\end{figure}

Simultaneous modelling \cite{SM} of the synchrotron and IC components 
constrains the magnetic field strength within a narrow 
interval,  {$B\simeq 112^{+15}_{-13} \mu$G} (see Fig.\ref{SED-onezone} 
and discussion below). 
The upper is set by the requirement for
the synchrotron radiation 
of the same electrons responsible for the production of 1~PeV photons 
to not overshoot the measured MeV flux \cite{Kuiper01}. 
The lower limit is set by requiring the
electron energy $E_{\rm e}$  and the accelerator's linear size, 
$l$, to meet 
the condition that the electron gyroradius $R_{\rm g}=E_{\rm e}/eB$  
($e$ is the charge of electron)  
cannot exceed $l$. Using Eq.(\ref{Ee}), we find 
\begin{equation}
(B/100 \mu {\rm G}) (l/1 \ {\rm pc}) \geq 0.023 (E_\gamma/1 \ \rm PeV)^{0.77} \ . 
\label{BL}
\end{equation}

Magnetohydrodynamic (MHD)  models of the Crab Nebula\cite{ReesGunn, KenCor} postulate 
that electrons are accelerated at the termination of a 
electron wind, then advected into the nebula through the MHD outflow. 
X-ray imaging of the inner parts of the nebula \cite{XrayImage} 
provides support for the location of the 
acceleration site being  close to the termination shock,  
at the distance $R \approx 0.1-0.14$~pc from the pulsar \cite{Meyer}.
The linear size of the accelerator should exceed the gyroradius, $l \geq R$;
this imposes a lower limit on the magnetic field from Eq.(\ref{BL}),  
$B \geq  20 \, \mu$G. 
On the other hand, the standard one-zone model, which assumes 
that both the synchrotron and IC components of radiation are produced
in the same region by the same electron population, gives 
$B \simeq 112 \,\mu$G (see Fig.\ref{SED-onezone}). Then,   
from Eq.(\ref{BL}) we find that 
$l \geq 0.02$~pc. These constraints are inconsistent with estimates of the characteristic size and magnetic field in the region(s) where  the flares  of the MeV/GeV $\gamma$-ray emission  (``Crab flares") \cite{Crabflares-agile} originate. The variation of $\gamma$-ray fluxes on timescales of days are 
interpreted by fast acceleration and synchrotron cooling of 
 PeV electrons in compact ($R\leq 0.01$~pc) highly magnetized ($B \geq 1$~mG)  regions 
(see \cite{CF-review} for a review). In the presence of such a large magnetic field,  the  IC $\gamma$-ray component is suppressed;  however this  does not exclude an indirect link between the PeV electrons responsible for the UHE $\gamma$-ray emission and the synchrotron  MeV/GeV flares.

The detection of $\sim 1$~PeV photons implies  an  acceleration rate which overcomes the 
synchrotron losses of the parent electrons 
up to PeV energies. 
The acceleration rate of electrons is
$\dot{E}=e \mathcal{E} c=\eta e B c$, where   
$\eta$ is the ratio of the projection of the electric field $\mathcal{E}$, averaged over the particle trajectory, to the magnetic field,  $\eta=\mathcal{E}/B$. This parameter  characterizing  the acceleration efficiency is always  smaller than 1; in objects where $\eta \to 1$, the accelerator proceeds at the maximum rate allowed by the classical electrodynamics and ideal MHD \cite{Synchmax}. 
The maximum energy of electrons then is determined by the balance between the acceleration and the energy {  loss} rates: $E_{\rm e, max}\approx 5.8 \eta^{1/2}(B/100 \,\mu \rm G)^{-1/2} \ \rm PeV$. 
Using the relation between 
$E_\gamma$ and $E_{\rm e}$ given by Eq.(\ref{Ee}), we find 
\begin{equation}
\eta=0.14 (B/100 \mu {\rm G}) 
(E_\gamma/1 \ \rm PeV)^{1.54} \ \rm PeV \ . 
\label{eta}
\end{equation}
Thus, for the detected $E_\gamma=1.1 \ \rm PeV$ and the magnetic field $B \simeq 112\,\mu$G derived from the one-zone model   (Fig.\ref{SED-onezone}), we find that the acceleration proceeds at the rate  $\eta \approx 0.16$.
For comparison, at the diffusive shock acceleration in  young supernova remnants\cite{MalkovDrury}, $\eta$ is smaller by at least 3 orders of magnitude. 

For the distance to the Crab, $d=2$~kpc \cite{CrabDistance}, the luminosity in PeV $\gamma$-rays is estimated
$L_{\rm \gamma,PeV}=4 \pi d^2 F_\gamma \approx 5 \times 10^{31} \ \rm erg \ s^{-1}$, where 
$F_\gamma \approx 10^{-13} \ \rm erg \ s^{-1}$ is the energy flux of  0.5 to 1.1~PeV $\gamma$-rays 
(Fig.\ref{SED-onezone}). In the Klein-Nishina limit, 
the IC cooling time of electrons in 2.7~K CMBR  
is $t_{\rm IC} \simeq 5 \times 10^{11} (E_{\rm e}/1 \ \rm PeV)^{0.7}\,$s \cite{Khangulyan14} 
or, for the given $E_\gamma$, using Eq.~\ref{Ee} we have $t_{\rm IC}\simeq 8\times 10^{11} (E_\gamma/1\,{\rm PeV})^{0.54}\,$s. 
Thus, the total energy held by the  $\geq 1$~PeV electrons responsible for production of $\geq 0.5$~PeV $\gamma$-rays is estimated as $W_{\rm e, PeV} = L_{\gamma, \rm PeV} t_{\rm IC} \approx 4 \times 10^{43} \ {\rm erg}$. Because the overall cooling of PeV electrons is dominated by synchrotron losses ($t_{\rm syn} \ll t_{\rm IC}$), the injection rate of PeV electrons 
$\dot{W}_{\rm e, PeV} = 
W_{\rm e, PeV} / t_{\rm syn} \approx 
2 \times 10^{36}(B/100 \mu \rm G)^2 \ \rm erg \ s^{-1}$. Thus,  within the framework of the standard  one-zone model with $B=112 \mu$G, the acceleration power of PeV electrons 
constitutes  $\approx 0.5 \, \%$ fraction of the pulsar's spin-down luminosity, $L_{\rm SD}\approx 5 \times 10^{38} \ \rm erg \ s^{-1}$\cite{KenCor}.  

\noindent{  \bf  Discussion}

We consider whether  the detection of PeV photons agrees with predictions 
for the standard MHD paradigm of 
the Crab Nebula which  assumes  that  nonthermal 
emission from X-rays to multi-TeV $\gamma$-rays \cite{dejager,AA96,Volpi2008,Meyer,Khangulyan2020} 
is produced by electrons 
accelerated at the termination of the pulsar wind.   
We modelled the Crab's multi-wavelength radiation within the
idealized Synchrotron-IC one-zone model, assuming a homogeneous spatial distribution of the magnetic field and electrons 
(Fig.~\ref{SED-onezone}). 
For $E_\gamma \geq 100$~TeV $\gamma$-rays 
the dominant  target for  IC scattering is  the 
2.7\,K CMBR, with properties that are known more precisely than the targets  at lower energies. 
For a steady-state electron energy distribution, above 1\,TeV,   we assumed 
a power-law function terminated by a super-exponential cutoff at the high-energy end.
Fig.~\ref{SED-onezone} demonstrates that using  three free parameters -  the power-law slope $\alpha=3.42$, cutoff energy $E_0=2.15\,$PeV and  magnetic field $B=112\, \mu$G, we reproduce the SED fit of reasonable accuracy from the
X-rays to multi-MeV $\gamma-$rays with synchrotron radiation, and the TeV to PeV $\gamma$-rays with IC radiation. 
Below 1\,TeV,   the electron spectrum must undergo a break to avoid a conflict with the synchrotron radiation at optical to radio frequencies, and
to provide a smooth transition of the IC radiation from TeV to GeV  energies \cite{SM}.  

The one-zone model in Fig.\ref{SED-onezone} is tightly constrained; 
the 3-sigma variation of the index $\alpha$ (3.37 to 3.47)  is determined
by the uncertainty of the X-ray data. 
Because $\geq 10$~TeV electrons are cooled on short timescales, 
the index of the initial (acceleration) spectrum must be close to 2.4.  
The magnetic field is determined by the flux ratio of the synchrotron and IC components, with 3 sigma range from 100\,$\mu$G to 130\,$\mu$G. For the given $\alpha$ and $B$, the cutoff energy $E_0$ is determined by the synchrotron and IC spectra
above 10~MeV and 100~TeV, respectively. The derived ranges of magnetic field  and the power-law index of electrons are consistent  with the previous studies based on 
the Synchrotron-IC one-zone model \cite{Meyer,Khangulyan2020} as well as the 
MHD flow \cite{AA96,Volpi2008,BAA2011} models in which the magnetic field's radial distribution is determined by the wind-magnetization parameter $\sigma_B$. The magnetic field {  $B\simeq 112$~G}
derived for the production region of multi-TeV to PeV $\gamma$-rays,  
is a factor of 2-3 smaller than the average nebular magnetic field, 
consistent with the MHD flow model\cite{KenCor}. The latter 
predicts  reduced  magnetic field at the 
termination shock for a broad range of the magnetization parameter $\sigma_B$ between 0.001 and 0.01 \cite{KenCor}. 

Within the one-zone model, 
the IC $\gamma$-ray spectrum is precisely predicted.  
While the KM2A spectral points from 10\,TeV to 1\,PeV  do agree, within 
the statistical uncertainties,  with the one-zone fit, there are   
possible deviations from its
predictions. Between 60 TeV and 500 TeV, two differ with a significance of  
$4 \sigma$ indicating a steeper spectrum than  the one-zone model predictions. The possible 
excess around one PeV indicates an opposite tendency - a hardening of the spectrum. A hardening of the electron spectrum is difficult to 
accommodate with plausible assumptions.  The problem of suppression of the one-zone spectrum at 1\,PeV can be circumvented by introducing a second population of PeV electrons. This could also explain the inconsistency of the synchrotron part of the SED with the one-zone model (Fig.\ref{fig:sed_2e})
by decoupling the highest energy Synchrotron and IC components, assuming that 
the MeV synchrotron radiation is predominantly 
produced in compact highly magnetized regions, while  the PeV IC photons originate from 
regions with $B \le  100 \mu \rm G$. 
A second electron component could  extend the SED to a few PeV but not 
much further. 
From  Eq.(\ref{eta}) follows  that,   
even for $\eta=1$ and 
minimum allowed magnetic field, $B \geq 20 \mu$G, the the maximum energy of photons 
cannot exceed 4~PeV. Any detection of $\gamma$-rays well beyond 1~PeV would require non-leptonic origin of the extra component of radiation, i.e. the presence of multi-PeV protons and nuclei in the nebula.  

Because of the limited energy budget available for acceleration of protons,  
hadronic interactions cannot be responsible for the overall 
broad-band $\gamma$-ray luminosity.  
Yet,   the 
contribution of hadronic interactions at PeV energies could be noticeable. 
%
The $\gamma$-ray production efficiency of hadronic interactions, i.e. the fraction of the proton kinetic energy converted to $\gamma$-rays, is determined by the ratio
$\kappa=t_{\rm esc}/t_{\rm pp}$, where $t_{\rm esc}$ is the 
confinement time of protons inside the nebula, and $t_{\rm pp} \approx 1.5 \times 10^{8} n^{-1} \ \rm yr $ is the cooling time of protons through the production 
and decay of the secondary $\pi^0$-mesons. For the average 
density of the nebular gas $n \approx 10\,\rm cm^{-3}$, the radiation efficiency is low;    
even for the most effective confinement of  10 PeV protons, 
$t_{\rm esc} \leq 250\,$yr \cite{SM}, thus $\kappa \leq 5 \times 10^{-5}$.  To explain 
the PeV $\gamma$-ray luminosity, $L_\gamma \approx 5 \times 10^{31} \ \rm erg \ s^{-1}$,
the acceleration power of $\sim 10$~PeV parent protons would need to be      
$\dot{W}_{\rm p}=W_{\rm p}/t_{\rm esc}=\kappa^{-1} L_\gamma \approx 10^{36} 
\ \rm erg \ s^{-1}$  or, in the case of  a broad $E^{-2}$ type proton spectrum,
an order of magnitude larger. These estimates are supported by numerical calculations presented in Fig.\ref{fig:sed_e+p}. They avoid  exceeding  
the theoretical constraints on the proton fraction \cite{Arons03} only 
if there is effective proton confinement in the  nebula.  


\newpage

\renewcommand{\refname}{References}
\bibliographystyle{science}
\bibliography{ms}

\setcounter{firstbib}{\value{enumiv}}

\noindent {\bf \large Acknowledgments} \\
This work is supported in China by National Key R\&D program of China under the grant  2018YFA0404201, 2018YFA0404202,  2018YFA0404203, 2018YFA0404204,  by NSFC (No.12022502, No.11905227, No.U1831208, No.U1931112, No.11635011, No.11761141001, No.11905240, No.11675204, No.11475190, No.U2031105, No.U1831129), in Thailand by RTA6280002 from Thailand Science Research and Innovation. The authors would like to thank all staff members who work at the LHAASO site above 4400 meter above the sea level year round to maintain the detector and keep the electricity power supply and other components of the experiment operating smoothly. We are grateful to Chengdu Management Committee of Tianfu New Area for the constant financial supports to the researches with LHAASO data.\\

\noindent {\bf \large Author Contributions}\\
Z.Cao and F.Aharonian contribute the major sections of the text about the experiment and interpretation, respectively. S.Z.Chen leads the KM2A data analysis with the team members C.Li and L.Y.Wang and others. LYW carried out event-by-event analysis and corresponding simulation. S.J.Lin and M.Zha lead teams to conduct the spectrum analysis using WCDA data and corresponding cross checking. S.S.Zhang leads the WFCTA team including L.Q.Yin to finish the combined analysis of all data from WFCTA, WCDA and KM2A for commonly registered events. LQY carries out the specific multi-component analysis and corresponding simulation. F.Aharonian and R.Y.Liu conduct the theoretic interpretation. RYL produces all corresponding calculations and all figures. He also contributes a lot for the manuscript edition. ZC is the spokesperson of LHAASO Collaboration and PI of the LHAASO project in China. He leads the specific working group for this paper involving all corresponding authors. Editorial Board of LHAASO collaboration led by S.M.Liu and D.della Volpe organizes internal review in the collaboration on the manuscript before submission and keeps informing the entire collaboration about updates and collects internal comments. B.D’Ettorre Piazzoli as the science advisor of the collaboration makes special contribution in manuscript revision and corresponding discussion. All other authors participate data analysis, including event reconstruction, simulation and event building with multi components, detector calibration, operation and maintenance of all scintillator counters, muon detectors, water ponds and Cherenkov telescopes. In the specific period of time, detector arrays are enlarged while the data were taken for this paper. Many authors participate the detector construction and deployment. All the contributions justify them to meet the Science’s authorship criteria.

\noindent {\bf \large Competing Interests}\\
There is no conflict of interest of the collaboration members. All relevant funding grants are listed in the Acknowledgments section.

\noindent {\bf \large Data and materials availability}\\
Data and software to reproduce these results are available on the LHAASO public data web page at http://english.ihep.cas.cn/doc/4035.html , including the event list, results of the background rate calculations, numerical values of the derived WCDA and KM2A spectra (Fig. 3) and SED (Fig. 4), and the code used to produce the SED and significance maps (Fig S1). The software we wrote for simulation of the detectors and background rate calculations is restricted by the terms of a grant from China's National Commission of Development and Reform due to legal restrictions imposed during the construction phase of LHAASO. Readers who are willing to be associated members of LHAASO Collaboration can request a copy under condition that any paper due to the corresponding analysis is the collaboration paper, signed by all authors including the associated ones. Interested readers can contact the corresponding authors for details.\\

\noindent {\bf \large Supplementary Materials}: LHAASO Collaboration authors and affiliations, Materials and Methods, Supplementary text, Figures S1-S9, Tables S1-S2, References (\textit{42-64})

\vfill
\clearpage

\begin{center}
{\LARGE Supplementary Materials for\\
\vspace{5pt}
PeV $\gamma$-rays from the Crab}
\end{center}

\vspace{100pt}

\noindent{\bf This PDF file includes}
\begin{itemize}
    \item[] Full list of LHAASO Collaboration authors and affiliations
    \item[] Materials and Methods
    \item[] Supplementary Text
    \item[] Figs. S1 to S9
    \item[] Tables S1 to S2
\end{itemize}

\clearpage

\setcounter{equation}{0}
\renewcommand{\theequation}{S\arabic{equation}}

\setcounter{figure}{0}
\renewcommand{\thefigure}{S\arabic{figure}}

\setcounter{table}{0}
\renewcommand{\thetable}{S\arabic{table}}

\section*{LHAASO Collaboration Authors and Affiliations}
\noindent
{Zhen Cao$^{1,2,3}$,
F. Aharonian$^{4,5}$,
Q. An$^{6,7}$,
Axikegu$^{8}$,
L.X. Bai$^{9}$,
Y.X. Bai$^{1,3}$,
Y.W. Bao$^{10}$,
D. Bastieri$^{11}$,
X.J. Bi$^{1,2,3}$,
Y.J. Bi$^{1,3}$
% ...
}\\

\footnotesize{
\noindent
$^{1}$ Key Laboratory of Particle Astrophyics \& Experimental Physics Division \& Computing Center, Institute of High Energy Physics, Chinese Academy of Sciences, 100049 Beijing, China\\
$^{2}$ University of Chinese Academy of Sciences, 100049 Beijing, China\\
$^{3}$ Tianfu Cosmic Ray Research Center, Chengdu, Sichuan,  China\\
$^{4}$ Dublin Institute for Advanced Studies, 31 Fitzwilliam Place, 2 Dublin, Ireland \\
$^{5}$ Max-Planck-Institut for Nuclear Physics, P.O. Box 103980, 69029  Heidelberg, Germany\\
$^{6}$ State Key Laboratory of Particle Detection and Electronics, China\\
$^{7}$ University of Science and Technology of China, 230026 Hefei, Anhui, China\\
$^{8}$ School of Physical Science and Technology \&  School of Information Science and Technology, Southwest Jiaotong University, 610031 Chengdu, Sichuan, China\\
$^{9}$ College of Physics, Sichuan University, 610065 Chengdu, Sichuan, China\\
$^{10}$ School of Astronomy and Space Science, Nanjing University, 210023 Nanjing, Jiangsu, China\\
$^{11}$ Center for Astrophysics, Guangzhou University, 510006 Guangzhou, Guangdong, China\\
% ...
}

\normalsize

\section*{Materials and Methods}
\vspace{5pt}

\section{Experiment Description}

LHAASO~\cite{Cao10} is a complex of
extensive air shower (EAS) detectors installed on 
Mt. Haizi (29$^\circ$21'27.6" N, 100$^\circ$08'19.6" E) at  4410 m 
above sea level, in Sichuan province, China.

\subsection{KM2A} 
KM2A consists of 5195 scintillator counters on the surface and 1188 underground  muon detectors spread out over a circular area of 1.3 km$^2$. 
The details of these detectors and their operation are described elsewhere\cite{KM2A-on-Crab}.

\subsection{Water Cherenkov Detector Array}
The Water Cherenkov Detector Array (WCDA) is composed of 3 large ponds, is one of the major components for the $\gamma$-ray astronomy in the LHAASO experiment.
The first pond is 150 m$\times$ 150 m (WCDA-1), equipped with 900 pairs of 8-inch and 1.5-inch photomultiplier tubes (PMT) arranged on a square grid with a side of 5 m at 4.4 m beneath the water surface. The detector, which is active everywhere in the whole area of the pond, collects  Cherenkov light generated by shower particles passing through the water. The total amount of Cherenkov photons are proportional to energies carried by the particles, except for the muons, traversing the water. Calibrated PMTs at the centers of $5\ \rm m\times5\ m$ cells measure both number of photons and their arrival time. The PMTs are synchronized within 0.2 ns. This enables the shower arrival direction to be measured with a resolution of 0.2$^\circ$ above a few TeV. Muons generate stronger signals than other particles given the long track they produce, being more penetrating  in water. Because electrons/positrons and $\gamma$-rays laterally extend in space following a steep distribution function, at certain distance, e.g. 45 m from the shower center, density of those particles drops very low and muon signals dominate, thus the absence of those hot spots in the shower footprints at far distance is used to indicate electromagnetic cascade showers induced by primary $\gamma$-rays. In this way, the cosmic ray background is suppressed at least  by a factor of 100. This provides WCDA-1 with a $\gamma$-ray source survey sensitivity of 65 milli-Crab with 5 $\sigma$ in a year, or equivalently, any source having an energy flux stronger than  $4\times 10^{-13}\, \rm erg/cm^{2}s$ above 3 TeV. { In 10 months of WCDA-1 observations, we detected the Crab Nebula with a significance of 77.4 $\sigma$}. The corresponding Spectral Energy Distribution (SED) of the Crab Nebula is determined by combining these data with previously published measurements at other wavelengths as shown in Fig. 2. More detailed description about the measurements of the SED is available elsewhere\cite{WCDA-on-Crab}.


\subsection{Wide FoV Cherenkov Telescope Array }
8 Cherenkov telescopes, each covering a field of view (FoV) of $16^\circ\times16^\circ$, have been operated for cosmic ray observation since October 2019. Showers  hitting the LHAASO array falling within the FoV of telescopes will be recorded simultaneously. To perform a coincident measurement, telescopes are located at the central region of the array, 6 installed at the south-west corner and 2 at the south-east corner of WCDA-1. Seven telescopes are set to cover a zenith angle range from 22$^\circ$ to 38$^\circ$, while one telescope is pointing toward zenith. For the showers detected by telescopes, in coincidence with WCDA  or KM2A,  the arrival direction is reconstructed with a precision of   0.2$^\circ$ and the core location on ground with an accuracy better than 3\,m for showers above 100 TeV. This allows the calculation of Cherenkov photons yield at various depth in the atmosphere. As long as the sky is clear, the Cherenkov light, mainly near ultraviolet photons, can be recorded no matter how high they were produced in atmosphere. Depending on the arrival direction of photons, an image is formed on a part of the  camera, which is installed at the center of the focal plane of the telescope and equipped with 1024 pixels.   Each pixel, made of $15\,{\rm mm}\times 15\,{\rm mm}$ silicon photomultipliers (SiPM), can measure the number of photons with linear response over a range from 10--30,000\cite{WFCTA-NIM}. Adding up the number of photons { over all pixels in the shower image is a sampling of the Cherenkov light pool, which is an accumulation of Cherenkov photons produced in the entire shower development. }
The brightness of the image is an estimator of the shower energy after correcting for a moderate effect on the distance between the telescope and the core location on the ground array, as the brightness decreases with the distance from the core. Then, the shower energy is reconstructed by establishing a response function between the total number of photons  and the primary energy of the incident particle using a Monte Carlo simulation which models the shower physics, its  development in the atmosphere, the Cherenkov photon production and propagation, but also the telescope responses as the photons enter in the instrument.

Finally, the reconstructed energies of showers are calibrated for the absolute energy scale. The calibration was performed by comparing energies with the ones reconstructed using WCDA on a group of commonly triggered events. The WCDA absolute calibration was performed  by measuring the deflection of the Moon shadows in the geomagnetic field in the energy range from 6 TeV to 40 TeV~\cite{energy-scale}.

\section{ {Observation of the two $\gamma$-ray events around 1\,PeV and SED of the Crab}}

\begin{figure}[H]
\centering\includegraphics[width=0.99\linewidth]{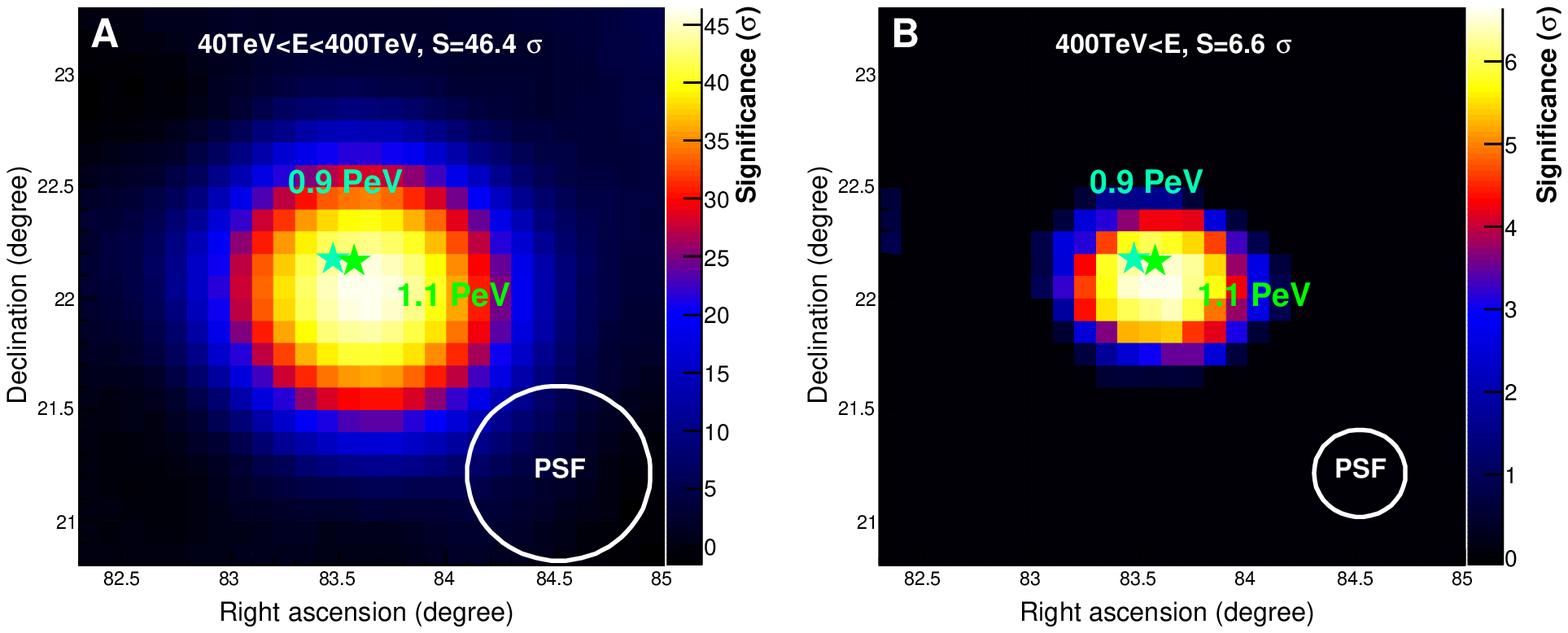}
\caption{{\bf Significance maps of Crab Nebula}. The map is centered on the Crab Nebula at energy 40$-$400 TeV (panel A) and $>$ 400 TeV  (panel B). The circles indicate the   PSF of KM2A. The color scale represents the significance. S is the maximum value in the map. The stars mark the direction of the 0.9\,PeV and 1.1\,PeV events.}
\label{fig:Crab-map}
\end{figure}

In the local coordinate system, the 0.88\,PeV event arrived at about 2 am on January 12th, 2020 (Beijing time), with a zenith angle of $(33.9\pm0.2)^\circ$ according to the shower front reconstructed by the 395 registered scintillator counters, which are synchronized within 0.2 ns. Fig.\ref{fig:Crab-map} show the significance map around Crab Nebula region. A clear signal at 6.6 $\sigma$ is observed at energies above 400 TeV. The 0.88 PeV event is found just $0.21^\circ$ away from the celestial coordinates of the Crab Nebula, as illustrated in Fig.\ref{fig:Crab-map}, consistent with the uncertainty in the reconstructed direction.

\begin{figure}
\centering\includegraphics[width=0.8\linewidth]{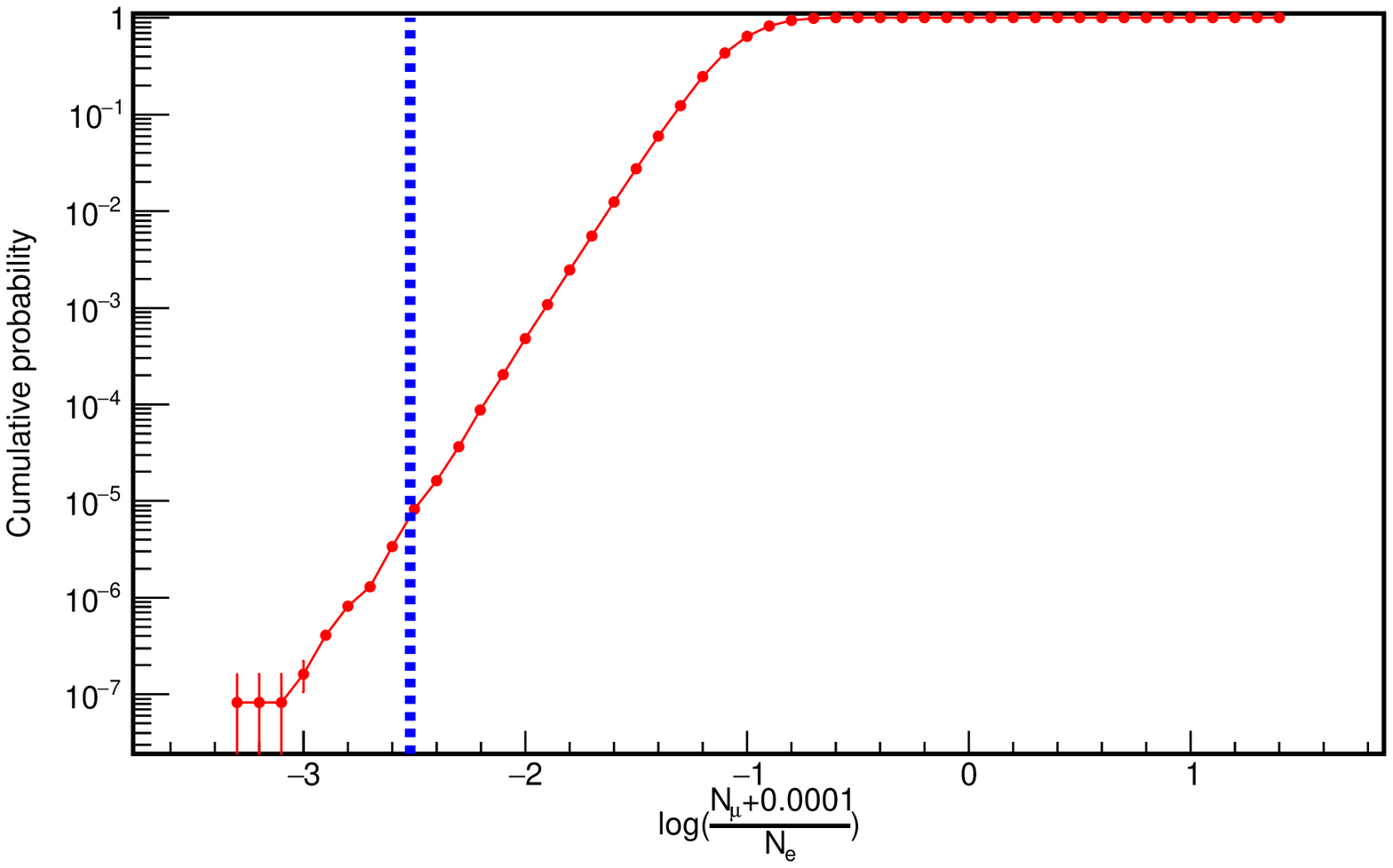}
\caption{{\bf Cumulative probability} of  $\log_{10}(({N_\mu+0.0001})/{N_e})$ for 12.3 million cosmic ray events detected $\pm 12^\circ$ away from the galactic plane in the first 308 days of data taking, with reconstructed energy E$_{\rm rec}>$0.88 PeV. The vertical dotted line indicates the value recorded in the 0.88 PeV event, i.e. -2.52. Only very few events are found below this value.}
\label{fig:CR-rejection}
\end{figure}

The event is only likely to be associated with the Crab Nebula if it can be identified as a $\gamma$-ray event. We used the muon detector (MD) array as a veto~\cite{KM2A-on-Crab}. The 15 muons ($N_\mu$) was compared to the total number of particles measured by all counters ($N_e$), 4996, which indicates a very small muon content in this event. The probability of being a hadron induced shower,
with such a small muon content, was evaluated by counting how many events have the ratio $N_\mu/N_e$ less than the ratio of 15/4996. It is found that only a few showers out of 12.3 million cosmic ray samples recorded by KM2A can have the reconstructed energy E$_{rec}\geq$0.88 PeV. We estimate the chance probability of any single cosmic ray event to be recognized as a $\gamma$-ray event as $6.7\times 10^{-6}$, as illustrated in Fig.\ref{fig:CR-rejection}. To avoid the pollution of diffuse $\gamma$-rays produced within the Milky Way, only events coming from regions 12$^\circ$ away from the Galactic plane are considered in the estimation. Taking into account that the Crab transits have specific occupancy of different zenith angles, this probability has been tested using events in 3 zenith angle intervals, i.e. 0$^\circ$-20$^\circ$, 20$^\circ$-40$^\circ$ and 20$^\circ$-50$^\circ$ and a negligible difference was found. 
In the direction of the Crab Nebula, KM2A recorded a total of 179 events with the angular distance from Crab Nebula less than 0.3$^\circ$, which is slightly larger than the radius of the PSF of KM2A above 400 TeV. Using the probability given above, the number of expected background CR events is 0.0012. In  other words, the chance probability of this event was not a photon is 0.1\%. 

\begin{figure}
\centering\includegraphics[width=0.8\linewidth]{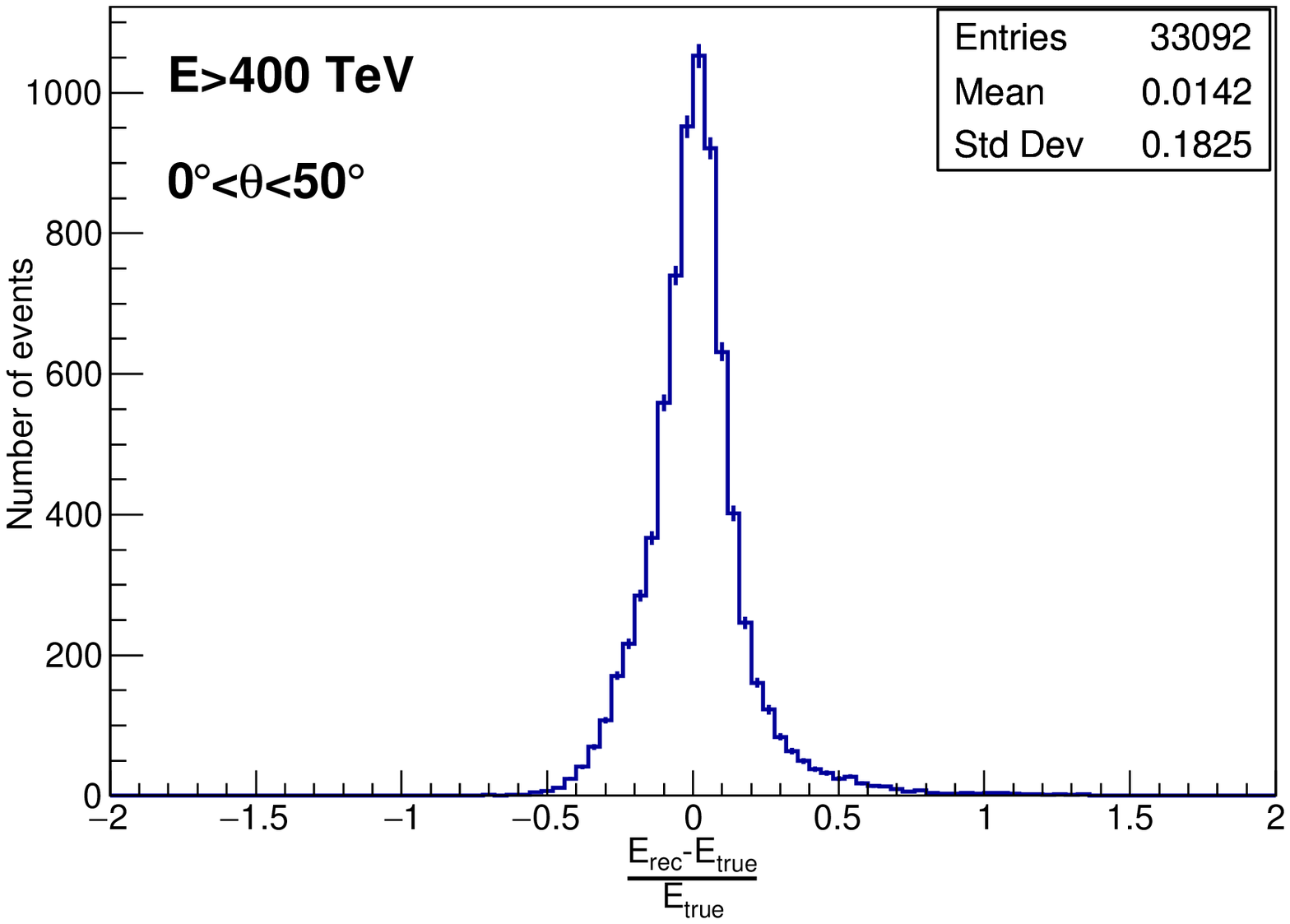}
\caption{{\bf The energy resolution of KM2A} for photon induced showers with incident zenith angle less than 50$^\circ$ and energy above 0.4 PeV. E$_{\rm rec}$ is the reconstructed energy and E$_{\rm true}$ is the input energy in the air shower and detector response simulation. The systematic shift is less than 1.5\% and the standard deviation is 18\%.}
\label{fig:E-res}
\end{figure}

The 395 scintillator counters measured the lateral distribution of the 4996 secondary particles in the shower. The core of this shower was reconstructed by fitting the distribution with the modified Nishimura-Kamata-Greisen (NKG) functional form, 
$N\Gamma(4.5-s)/\Gamma(s-0.5)/\Gamma(5-2s)(r/R_M)^{s-2.5}(1+r/R_M)^{s-4.5}$, where $N$ is the total number of secondary particles in the shower, $s$ is the shower age parameter, and $R_M$ is the Moliere unit of 136 m\cite{MU} as a measure of the width of the distribution at the altitude of LHAASO, and $r$ is the perpendicular distance from the detector to the shower axis. The uncertainty of the core location is about 3 m.  Here, the particles density at 50 m from the core determined in the fitting has been used as the energy estimator and has resulted in the shower energy  $E_{rec}=0.88\pm0.11$ PeV by assuming the primary particle a $\gamma$-ray photon. As a pure electromagnetic cascade shower, the energy scale and the resolution established using existing simulation tools for air showers and detector responses. The resolution defined as the $(E_{rec}-E_{true})/E_{true}$, where $E_{true}$ is the input shower energy in the simulation, is found to have a Gaussian form with the systematical shift less than 1\% and a resolution of 14\% for showers above 100 TeV and arriving with angles less than 20$^\circ$ from the zenith\cite{KM2A-on-Crab}. The equivalent values for showers above 400 TeV and from a zenith angle less than $50^\circ$ are the systematic shift $\sim 1\%$ and the resolution of $\approx 18\%$, see Fig.\ref{fig:E-res}.

To verify the estimation of the energy and its uncertainty of the 0.88 PeV photon, 10,000 events were simulated by using the geometric parameters of the measured event over a wide energy range. The distribution of input energies of events that have the total number of particles recorded by the registered scintillator counters and number of the counters are within $\pm$10\% around the detected values, is used to estimate the photon energy and its error as ($0.866^{+0.103}_{-0.097}$) PeV, in a good agreement with the reconstructed energy.


One of the WFCTA telescopes, Telescope No.10, pointing at a zenith angle of 30$^\circ$ and nearly toward west, saw the shower image in the lower 1/4 of its FoV starting from the zenith angle of 34$^\circ$.  Since the shower is quite far from the telescope, the image was not very much extended, only 11 pixels had signals above the threshold. However, the image is bright in terms of total number of photoelectrons, which was 8636. 30,000 $\gamma$-ray induced showers with the same geometric parameters were simulated in the energy range from 0.4 to 4 PeV. The generated showers reproduced the footprint on KM2A, WCDA and the image in the telescope. The most probable
energy among the simulated showers with $8636\pm100$ photoelectrons in their images  is found to be $0.92^{+0.28}_{-0.20}$ PeV.  

This same analysis was applied on all 5 events above 0.4 PeV. The results are summarized in Table \ref{tab:5-events}.
\begin{table}[htb]
  \centering
  \small
  \begin{tabular}{lcccccccccc}
    \hline
      E (PeV)& $\delta$E (PeV) & N$_e$ & N$_\mu$ & $\theta$ ($^\circ$) & D$_{\rm edge}$ (m)
      & $\psi$  ($^\circ$) & P (\%) &  {Arrival time}\\
    \hline \hline
1.12 & 0.09& 5094 & 14 & 13.0 & 89 & 0.15 & 0.03 & 2021-01-04 16:45:06\\
0.88 & 0.11& 4996 & 15 & 33.9 & 139 & 0.21 & 0.1 & 2020-01-11 17:59:18\\
0.57 & 0.13& 2408 & 9 & 40.8 & 125 & 0.08 & 0.7 & 2020-05-22 03:54:56\\
0.46 & 0.05& 2432 & 6 & 21.7 & 52 & 0.11 & 0.3 & 2020-11-05 21:23:28\\
0.40 & 0.04& 1859 & 3 & 23.1 & 65 & 0.10 & 0.2 & 2020-04-30 09:57:54\\
   \hline
  \end{tabular}
  \caption{{\bf Derived parameters for the 5 highest energy photons detected from the direction of the Crab.} E and  $\delta$E are the reconstructed photon energy and its error, respectively, in PeV. N$_e$ and N$_\mu$ are detected numbers of the secondary charged particles and muons in showers induced by the $\gamma$-ray photons, respectively. $\theta$ is the incident zenith angle of the shower in degrees. D$_{\rm edge}$ is the distance of the shower core from the nearest edge of the active detector array in meters. The larger distance indicates the better completeness of the shower detection. The typical radius of the densest area in a photon induced shower at 1 PeV is about 120 meter. $\psi$ is the space angle between the $\gamma$-ray event and the direction of the Crab in degrees. P is the percentage probability of the event being misidentified as a $\gamma$-ray induced shower.}
  \label{tab:5-events}
 \end{table}

The SED of Crab Nebula below 20 TeV was measured using data collected by WCDA-1 from  September  2019 to  {October} 2020.
The total exposure of the Crab Nebula in this period of time was  {343.5 transits of the Crab}. More details about the  {performances of WCDA} have been reported elsewhere\cite{WCDA-on-Crab}.  The independent data set collected in the operation of half KM2A for 314 days and three-quarter KM2A for 87 days from December 2019 to February 2021
has been used to measure the SED of the Crab Nebula above 10 TeV. More details about the measurement and detector performance have been reported elsewhere\cite{KM2A-on-Crab} for the half KM2A. Selected $>0.1\,$PeV $\gamma$-ray events according to the ratio $N_\mu/{N_e}<1/230$ were selected in a cone of 0.4$^\circ$ centered at the Crab Nebula direction. The criteria ensure that 90\% of signal is  contained, giving the angular resolution better than 0.25$^\circ$ of KM2A and the extension less than 0.07$^\circ$ for the Crab Nebula. The background cosmic rays are suppressed to be in the `signal-dominated' regime. In Table \ref{tab:Ns-Nb}, the details about the number of  on source photons and  off source  background events in the energy bins are listed. 

The energy resolution of KM2A is found to be about 18\% for all events with the zenith angle up to 50$^\circ$ as shown in Figure \ref{fig:E-res} for photon showers above 0.4\,PeV. A similar resolution function for lower energies has been reported \cite{KM2A-on-Crab}. Only a small non-Gaussian tails less than 1\% is found in the range of 0.5 PeV and above. The systematic shift is less that 1.5\%. Breaking into zenith angle ranges, the resolution is found better than 15\% for events with the zenith angle less than 40$^\circ$ and 30\% for events in the range of $40^\circ<\theta <50^\circ$. Given the resolution, the energy is divided into 5 bins per decade with a bin width of $\Delta log_{10}E =0.2$ to ensure the bin-to-bin migration mainly happens in the adjacent bins according to the Gaussian-shaped resolution function. In the operation of KM2A, the numbers of the events in bins above 0.1 PeV are listed in Table \ref{tab:Ns-Nb}. The number of background events N$_b$ in each bin is estimated by using the direct-integration method. The flux is measured for each bin as a function of $E$, as presented in Fig.\ref{fig:Crab-SED}. Many more technical details of the SED measurements can be found elsewhere\cite{KM2A-on-Crab}.

Above 400 TeV, the Crab Nebula has been detected with the significance of 6.6 $\sigma$. The sky map around the Crab is shown in Fig. \ref{fig:Crab-map}.

\begin{table}[htb]
  \centering
  \small
  \begin{tabular}{lccc|ccc}
    \hline
 ~ & $\theta < 50^\circ$ & ~ & ~ & ~ & $40^\circ < \theta < 50^\circ$ & ~ \\
    \hline
      log$_{10}(E/{\rm TeV})$ & N$_{on}$ & N$_b$ & N$_s$ &  ~~N$_{on}$ & N$_b$ & N$_s$ \\
    \hline \hline
2.0-2.2 & 55 & 1.413 & 53.59 &  ~~8 & 0.55 & 7.45  \\
2.2-2.4 & 23 & 0.459 & 22.54 &  ~~1 & 0.10 & 0.90  \\
2.4-2.6 &  6 & 0.176 & 5.824 &  ~~0 & 0.00 & 0.00  \\
2.6-2.8 &  3 & 0.045 & 2.955 &  ~~1 & 0.00 & 1.00  \\
2.8-3.0 &  1 & 0.008 & 0.992 &  ~~0 & 0.00 & 0.00 \\
3.0-3.2 &  1 & 0.000 & 1.000 &  ~~0 & 0.00 & 0.00  \\
   \hline
  \end{tabular}
\caption{{\bf The numbers of detected on-source photons N$_{on}$, remaining background events N$_b$ and signals N$_s$ in 6 bins of reconstructed energy E above 0.1 PeV.} All events used in the SED analysis up to zenith angle of 50$^\circ$ are listed in the left part of the table. Those with the zenith angle greater than 40$^\circ$ are in the right part. Only one event in the bin 2.6-2.8 has a zenith angle less than 41$^\circ$ as listed in Table \ref{tab:5-events}. }
\label{tab:Ns-Nb}
 \end{table}

\section{Spectral Fitting}
{\bf The standard one-zone scenario} 
We mainly  focus on the high-energy part of the electron spectrum  which account for keV -- MeV and TeV -- PeV data, where a power-law type electron spectrum with a high-energy cutoff can match the SED. To avoid violating the radio -- optical and GeV data, which should arise from low-energy electrons from a much larger volume of the nebula, a low-energy spectral cutoff or break is needed. Such a low-energy cutoff or break could naturally correspond to the bulk Lorentz factor of the cold pulsar wind just before the termination shock, which for the Crab Nebula  {  is expected around $10^6$\cite{KenCor}}.
If we choose the break energy and the low-energy spectral index, we could also fit the low-energy emission phenomenologically. We here assume a broken-power-law function with a super-exponential high-energy cutoff for the electron spectrum, i.e, 
\begin{equation}
    \frac{dN}{dE_e}=N_0E_e^{-\alpha}\left[1+(E_e/E_b)^{-\Delta \alpha}\right]^{-1}\exp\left[-(E_e/E_0)^2\right]
\end{equation}
where $\alpha$ is the spectral index after the spectral break, and  {$\Delta \alpha$ is the spectral difference before and after the break}, $E_b$ is the break energy, $E_0$ is the cutoff energy. 
{  The parameter $\alpha$ 
should match simultaneously the X-ray and multi-TeV $\gamma$-ray spectra; it should be  confined 
within a narrow range, close to
$\alpha =3.4$, The parameters $\Delta \alpha$ and $E_b$ are important for fitting the low energy IC $\gamma$-ray and optical to radio synchrotron data, but are not directly linked  to UHE energies. For any reasonable parameters set, 
 {$\Delta \alpha \leq 2$ and $E_b \leq 1$~TeV}.} 
$N_0$ is the normalization factor which can be determined once the total energy in the electron spectrum $W_e$ is given, via $W_e=\int E_e\frac{dN}{dE_e}dE_e$. 

In the magnetic and radiation fields, electrons are radiatively cooled through the synchrotron radiation and the inverse Compton (IC) scattering. For the synchrotron radiation, we consider a turbulent magnetic field with Gaussian distribution of an average strength $B$ \cite{Derishev19}. For the IC radiation, we employ the 
2.7~K CMBR, the interstellar radiation field~\cite{Popescu17}, the far-infrared radiation excess from the nebula with the  temperature of 70\,K and energy density of $0.5\rm \, eV~cm^{-3}$. The synchrotron radiation of the same population of electrons is also considered in order to calculate the synchrotron-self Compton process. The synchrotron radiation is assumed homogeneously generated within the nebula of a radius of 1.8\,pc, and the volume-averaged photon density is enhanced by a factor of 2.24 with respect to the boundary of the nebula~\cite{AA96}. We then explore the parameter space evaluating the goodness of fit by the Pearson's chi-squared test with the data above 1\,keV. The best-fitting result (shown in Fig.~\ref{fig:Crab-SED}) gives $\chi^2/dof=158.4/99$, among which the 11 data points of KM2A contributes a value of  {36.9} corresponding to a $4\sigma$ deviation between the model prediction and the data. 

The comparison between the broadband SED and the best-fitting model is shown in Fig.~\ref{fig:Crab-RtoG}, where $\Delta \alpha=1.76$ and $E_0=0.76\,$TeV is employed. In the same figure, we show the contributions of 
electrons from different energy intervals to the SED at different energies.
For the magnetic field $B=112$~G,
the relation between the characteristic energies of the synchrotron and IC photons from the same electron, is shown in Fig.~\ref{fig:syn-ic}. For the IC scattering on an isotropic target photon field of blackbody/greybody radiation with temperature $T$, the relation between the electron energy and the mean energy of up-scattered photons $\bar{E}_\gamma$ is  described by~\cite{Khangulyan14}:
\begin{equation}
\frac{\bar{E}_\gamma}{E_e}=
\frac{4b}{4b + 0.3}\frac{\log(1+b)}{\log(1+4b)},   
\end{equation}
where $b=E_e kT$,  $k$ is the Boltzmann constant.
All energies in this formula are expressed in unit of $m_ec^2$. 

For $2.7$~K CMBR, a 2.3\,PeV electron is required to produce 1.1\,PeV photon as shown in Fig.~\ref{fig:e-to-g}. The characteristic energy of the synchrotron radiation is given by $\varepsilon_{\rm syn}=0.29\cdot 3E_e^2eB/4\pi m_e^3c^5$.

\section*{Supplementary Text}
\vspace{5pt}
\section*{Alternative Two-zone Scenarios} 
\noindent{\bf Two-component leptonic scenario}
In the spectral fitting of the KM2A data in the standard one-zone scenario, we find that the model overproduces the measured flux between 60 -- 500\,TeV, and it fails to reproduce the spectral flattening above 500\,TeV although the latter is not statistically significant at the present time. Such a model deficiency can be circumvented by introducing a second population of PeV electrons, accelerated presumably in different sites by different mechanisms \cite{Khangulyan2020, Lyutikov2019}.
We discuss two type of spectra for the second electron population. 
First, we assume a Maxwellian-type electron component, which may be accelerated by the magnetic reconnection process with a moderate magnetization ($10^{-3}< \sigma_B < 0.1$) \cite{Sironi13}:
\begin{equation}
\label{N2max}
    \frac{{\rm d}N_2}{{\rm d}E_e}=N_2E_e^2\exp\left(-E_e/E_{2}\right)
\end{equation}
where $N_2$ is determined by the given total energy budget $W_{e,2}$ for this component, and {  $E_{2}$ is the cutoff energy of the second-component electron spectrum.} 
We allow the magnetic field in the region of radiation produced 
by the second electron population, $B_2$, to differ from the field in the region occupied by the first electron population. 
The acceleration by magnetic reconnection in a medium 
with high magnetization ($\sigma_B > 10$), forms a steeper particle spectrum, e.g. 
a power-law spectrum with a slope $\sim 1.5$ \cite{Sironi14, Lyutikov2019}. Therefore, 
for  the second electron population we consider also the following spectrum:
\begin{equation}
\label{N2pl}
       \frac{{\rm d}N_2}{{\rm d}E_e}=N_2E_e^{-1.5}\exp\left(-E_e/E_{2}\right). 
\end{equation}
{  Because of different signs of the power-law index in Eqs(\ref{N2max}) and (\ref{N2pl}), 
the spectral fits of the same radiation data,  require different  cutoff energies $E_2$. For the  Maxwellian-type distribution, $E_2$ is significantly smaller than for the power-law distribution.}

\begin{figure}
    \centering
    \includegraphics[width=0.8\textwidth]{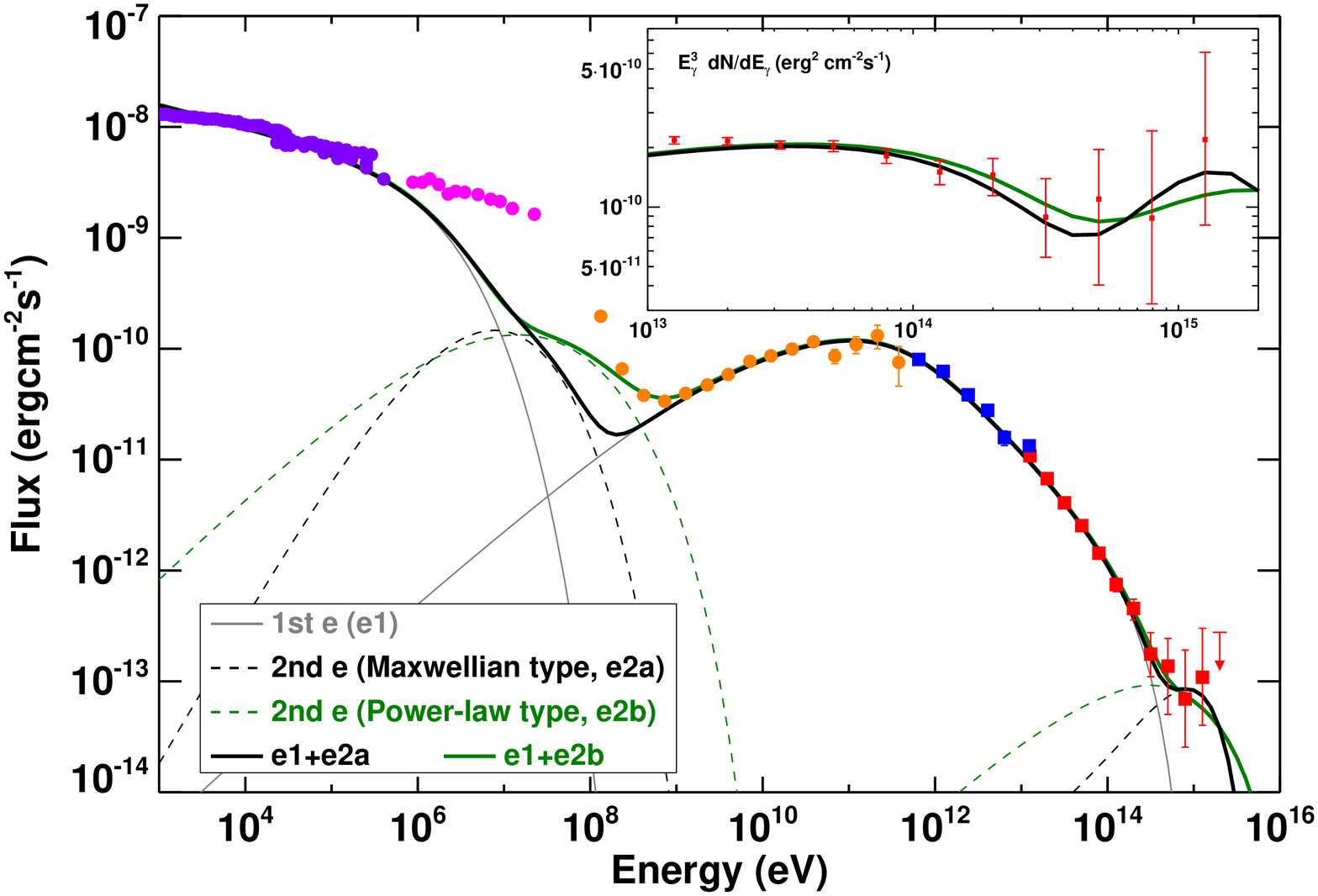}
    \caption{{\bf A two-zone scenario with two electron populations. }The COMPTEL data is assumed to arise from the flare state and is ignored in the fitting. The first electron component (e1) is the same as the one in Fig.~\ref{SED-onezone} except for the smaller cutoff energy $E_0=450\,$TeV. The second electron component is assumed to radiate in a weaker magnetic field $B_2=30\,\mu$G. Two type of  distributions are assumed for the 2nd electron population.
    (a)  Maxwellian-type spectrum of the "temperature" $E_2=400\,$TeV and  total energy $W_{e,2}=7\times10^{43}\,$ergs for $E_e>1\,$TeV (e2a). (b)  Power-law spectrum with a slope fixed at 1.5, an exponential cutoff at $E_2=2\,$PeV; the total energy  $W_e=1.5\times10^{44}\,$ergs (e2b). The solid grey curve shows the radiation of the first component. The dashed black and green curves represent, respectively the radiations of the second component in two spectral forms. The thick solid black curve shows the sum of e1 and e2a, while the thick solid green curve shows sum of e1 and e2b.}
    \label{fig:sed_2e}
\end{figure}

The synchrotron cooling limits the characteristics distance at which the highest energy electrons release all their energy: $l_{\rm rad} \leq 6 \times 10^{17} (E_e /1\,{\rm PeV})^{-1}(B/100\mu {\rm G})^{-2}\,$cm. The multiwavelength modelling of the synchrotron
and IC components of radiation does not allow a deviation of the average magnetic field from $100\,\mu$G implying that both the acceleration and radiation of PeV electrons take place in a compact region(s), attached, presumably, to the wind termination site. The short lifetime of electrons and their confinement in compact regions
allow flux variability in both channels - IC $\gamma$-rays at PeV and synchrotron radiation at MeV energies on timescales of months. The synchrotron radiation could  
experience variability on shorter timescales caused, e.g. due to the synchrotron burning of electrons in locations with suddenly enhanced, up to 1\,mG magnetic field, initiating  the so-called Crab flares.
Therefore, the MeV emission of the Crab Nebula measured previously \cite{Kuiper01} may originate from a region which is different from the production region of PeV photons. It will relax the procedure of the spectral fitting given additional model parameters, and the result is shown in Fig.~\ref{fig:sed_2e}. 

\noindent{\bf Lepto-hadronic scenario}
The production of hadronic $\gamma$-rays in the Crab Nebula is less likely but cannot be 
excluded \cite{AA96, ZhangX20}. 
Several locations in the Crab have been discussed in the literature 
as potential proton and nuclei acceleration sites.
One of them is the pulsar's magnetosphere where the fast magnetic 
rotator produces a potential difference between the pole and the equator where
protons can be accelerated if they experience a fraction of the voltage drop \cite{Blasi00, Arons03, Guepin20}. For the typical parameters of the Crab's pulsar, 
protons accelerated in this way may reach the energy of 30\,PeV \cite{Blasi00, Arons03} or even higher \cite{Guepin20}. At high energies, the time integrated spectrum may form a $E_p^{-2}$ type power-law spectrum \cite{Arons03}. 

Another possible acceleration site is the wind termination, through the relativistic shock or shock-driven magnetic reconnection \cite{Sironi13, Lemoine15}. The formed spectrum is 
expected to be power law with a slope between 2 and 3. Alternatively, hadronic emission could be produced by protons 
from the cold, ultrarelativistic pulsar wind with Lorentz factor $\Gamma_{\rm w}$ \cite{Amato03}. 
In this scenario, the protons are monoenergetic, $E_p=m_{\rm p} c^2 \Gamma_{\rm w}$.
To produce 1\,PeV $\gamma$-ray photon, the bulk Lorentz factor of the pulsar wind should 
exceed  {  $10^7$.  The latter is significantly larger than the ``standard'', for the Crab pulsar, 
value of $10^6$\cite{KenCor},  although cannot be  excluded for the 
small $e^{\pm}$ electron-positron  multiplicity $k_\pm  \sim 10^3$ \cite{Arons2008}.}   

\begin{figure}
    \centering
    \includegraphics[width=0.8\textwidth]{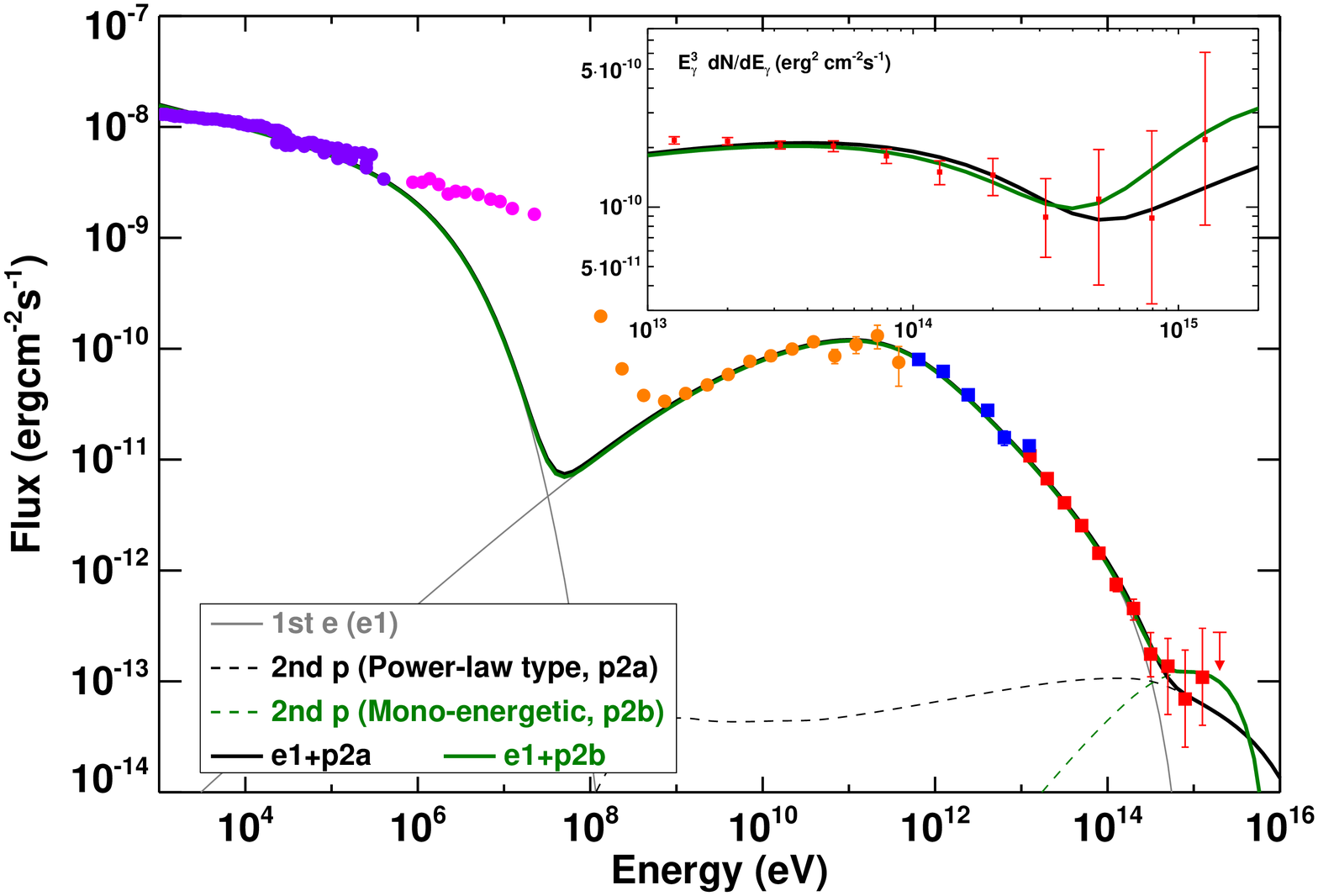}
    \caption{{\bf A two-zone scenario with the the main (electron) and second (proton)
    relativistic particle populations.} The first electron component (e1) is the same as the one in Fig.~\ref{fig:sed_2e}. The dashed black and green curves represent the $\pi^0$-decay $\gamma$-ray flux calculated for (a) the  proton spectrum of the form $E_p^{-2}\exp(-E_p/{30\rm \,PeV})$ with the total energy $W_p=3\times 10^{47}\,$erg (p2a) 
    and (b)  monoenergetic 10~PeV protons with $W_p=5\times 10^{46}\,$erg (p2b,) respectively. The thick solid black curve represents the sum of e1 and p2a, while the solid thick green curve represents the sum of e1 and p2b, respectively.}
    \label{fig:sed_e+p}
\end{figure}

\begin{figure}
  \centering\includegraphics[width=0.8\linewidth]
  {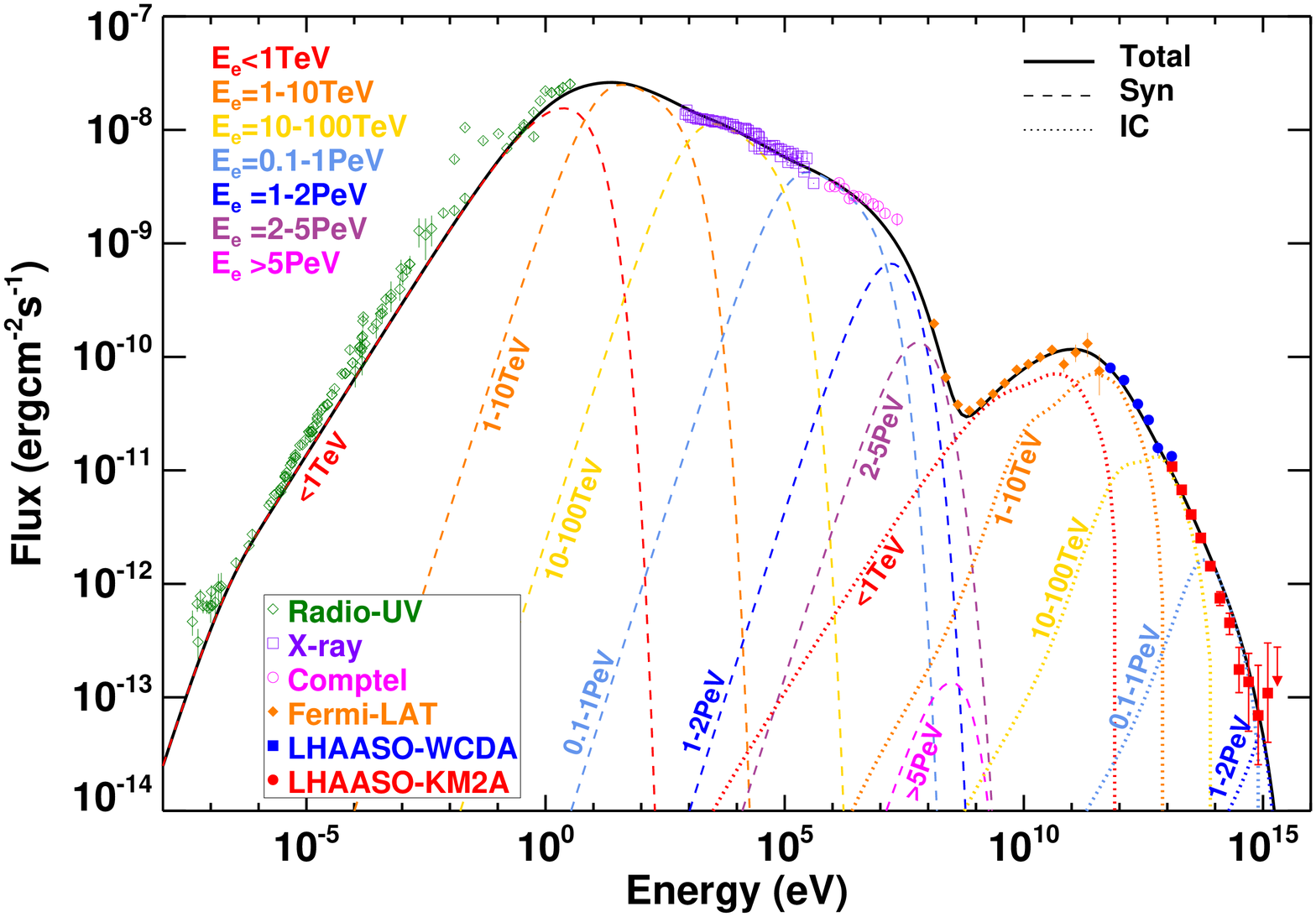}
  \caption{{\bf Contribution of electrons to 
  different energy bands of the SED of the Crab Nebula from radio to UHE $\gamma$-rays}. All the parameters are same with that in Fig.~\ref{SED-onezone}  {with additionally introducing the spectral break at $E_c=0.76\,$TeV in the electron spectrum and a hardening of the spectrum by $\Delta \alpha=1.76$ before the break.}
  The multi-wavelength observations of the Crab Nebula are compared with theoretical 
  expectations calculated within the one-zone model. The calculated fluxes are decomposed to demonstrate the contributions of electrons in the radiation in different energy bands.  The dashed and dotted curves show the synchrotron and  IC radiation components with the best-fit parameters. The the black solid curve is the sum. The open green diamonds show the radio\protect\cite{Baldwin71,Macias-Perez10}, infrared \protect\cite{Ney68,Grasdalen79,Green04,Temim06},
  optical \protect\cite{Veron-Cetty93}, and ultra-violet\protect\cite{Hennessy92} measurements. The open purple squares and open magenta circles represent  the compiled X-ray  and soft $\gamma$-ray (COMPTEL) fluxes \protect\cite{Kuiper01}, respectively.  The filled orange diamonds represent the Fermi-LAT data for the non-flare state\protect\cite{Arakawa20}. Filled blue squares and filled red circles are the LHAASO (WCDA and KM2A, respectively) measurements.} 
  \label{fig:Crab-RtoG}
\end{figure}

\begin{figure}
    \centering
    \includegraphics[width=0.8\textwidth]{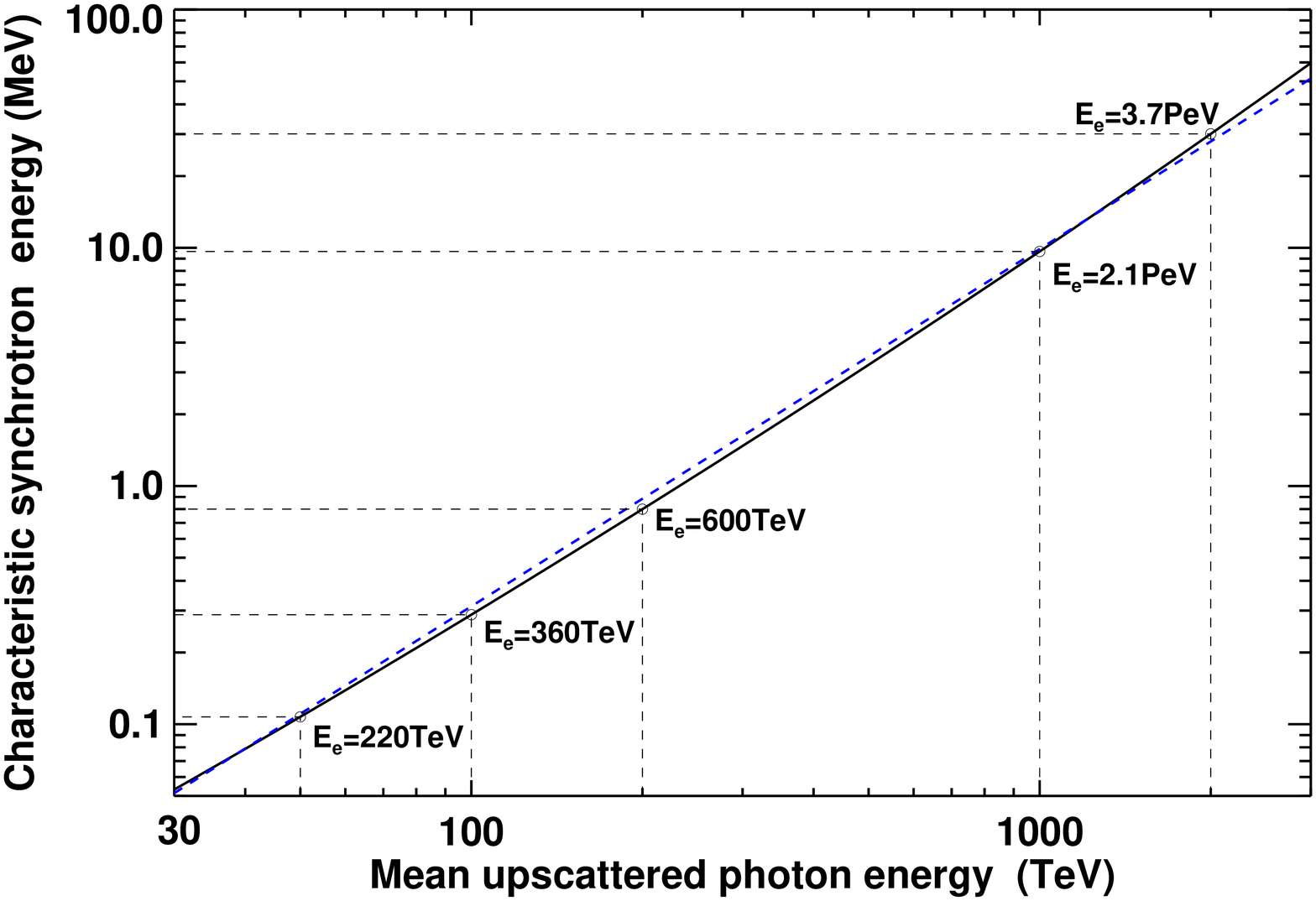}
    \caption{{\bf Relation between the characteristic energies of the upscattered  (from 2.7K CMBR) IC photon and the  the synchrotron photon  radiated by the same electron for the magnetic field $B=112$~G.} The solid line shows the numerical result while dashed line corresponds to the approximate analytic formula $\varepsilon_{\rm syn}=10(\bar{E}_{\gamma}/1{\rm PeV})^{1.5}\,$MeV which within the interval $30\,{\rm TeV}\leq \bar{E}_{\gamma} \leq 3\,{\rm PeV}$ provides an accuracy better than 15 \%.   The open circles on the solid line indicate energies of parent electrons emitting 50\,TeV, 100\,TeV, 200\,TeV, 1\,PeV, and 2\,PeV $\gamma$-ray photons.}
    \label{fig:syn-ic}
\end{figure}

\begin{figure}
\centering\includegraphics[width=0.8\linewidth]
{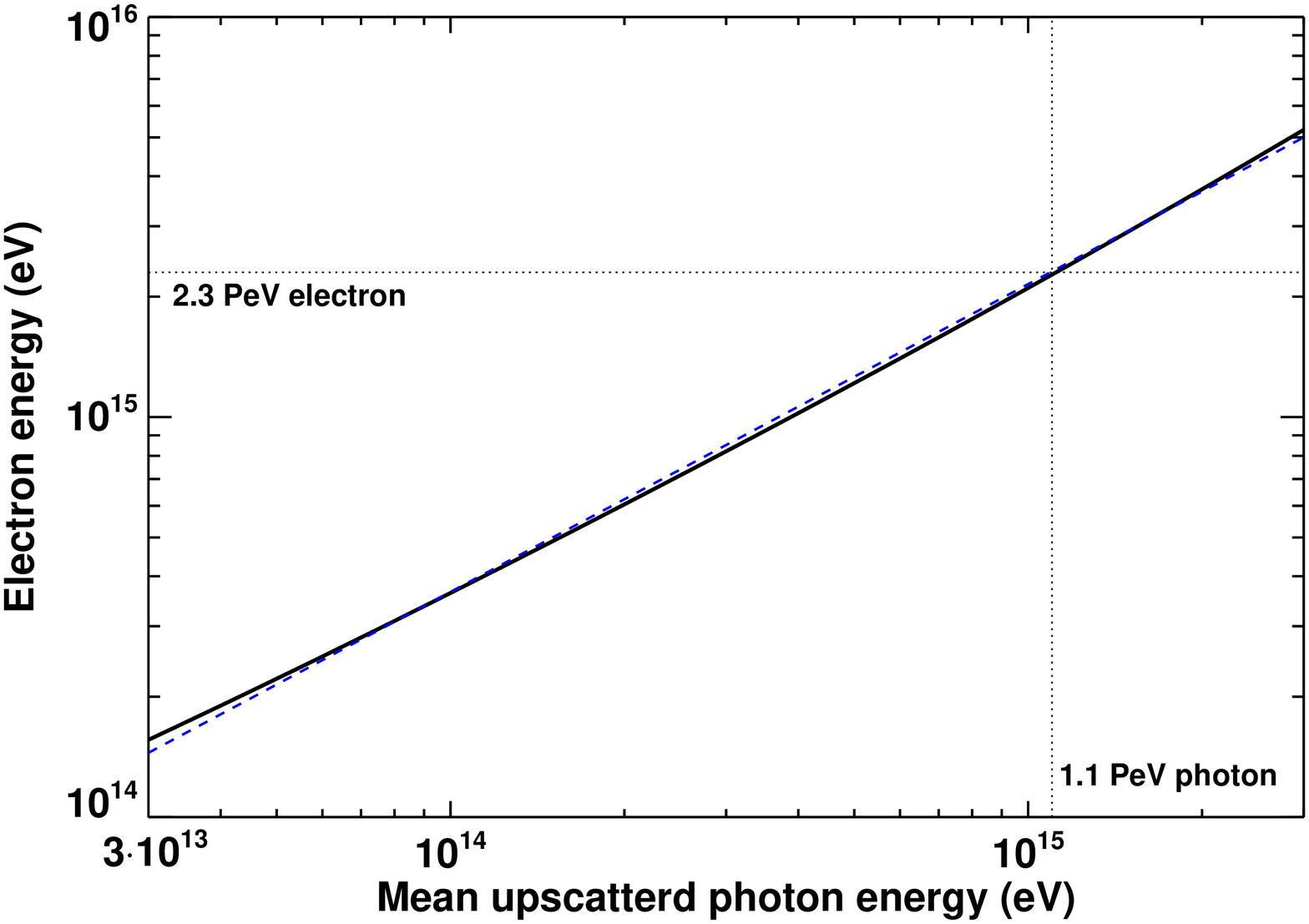}
\caption{{\bf Relation between the average energy of upscattered IC photons (from 2.7K CMBR) and the energy of the parent  electrons.} The solid line shows the numerical result while the dashed line represents an analytic approximation $\bar{E}_{\rm IC}=0.37(E_e/{\rm 1PeV})^{1.3}$, which within the interval $30\,{\rm TeV}\leq \bar{E}_{\rm IC} \leq 3\,{\rm PeV}$ provides an accuracy better than 10\% .} 
\label{fig:e-to-g}
\end{figure}

In Fig.~\ref{fig:sed_e+p}, we consider two types  of proton spectra: 
\begin{equation}
\frac{dN_{\rm p}}{dE_{\rm p}}=N_{\rm 0,p} E_{\rm p}^{-2} \exp(-E_{\rm p}/{\rm 30\, \rm PeV}),
\end{equation}
and
\begin{equation}
    \frac{dN_{\rm p}}{dE_{\rm p}}=N_{\rm 0,p}\delta(E_{\rm p}-{\rm 10\,PeV})
\end{equation}
The normalization factor $N_{\rm 0,p}$ determines   the total energy of protons.  
The target density for the hadronic interactions 
can be calculated from the the  total mass of gas and dust in the nebula,
$M=7.2\,M_\odot$ with $M_\odot$ being the solar mass \cite{Owen15}.  For the radius of the Crab Nebula,  
$R=1.8$\,pc,  the mean number density of nucleons (independent of the 
gas composition) is estimated $10\,\rm cm^{-3}$. 
To calculate the spectra of $\gamma$-rays, we use a semi-analytical method \cite{Kafexhiu14}.
The results are  shown in Fig.~\ref{fig:sed_e+p}. 

The total energy in protons 
to explain the $\gamma$-ray flux at 1 PeV is 
$3 \times 10^{47}\,$erg
for the power-law distribution of protons,   
and $5\times10^{46}\,$erg for  monoenergetic 10~PeV protons. 
The particle escape time in the nebula determines the acceleration power of protons,
$\dot{W}_{\rm p}=W_{\rm p}/t_{\rm esc}$. 
We consider two extreme cases: (i) protons escape ballistcally  with  the speed of light and (ii) propagate diffusively in the Bohm limit. The former case corresponds to the fastest escape and the timescale is comparable to the light crossing time of the entire nebula, i.e., $t_{\rm esc}=R_{\rm pwn}/c\simeq 6$\,years. In the latter case, for the mean magnetic field in the nebula  $B=300\,\mu$G, the Bohm diffusion coefficient is $D_B\simeq 10^{27} (E_p/10\,{\rm PeV})(B/300\,\mu \rm G)^{-1}\rm cm^2s^{-1}$. The diffusive escape time  is estimated as 
$t_{\rm esc}=R_{\rm pwn}^2/4D_B\simeq 250 (E_p/10\,{\rm PeV})^{-1}(B/300\,\mu \rm G)$\,years. 
Correspondingly,  the luminosity of 10~PeV protons 
estimated  in the broad limits $\dot{W_{\rm p}}=W_{\rm p}/t_{\rm esc} \sim (7-260)\times 10^{36} (W_p/5\times10^{46}\rm \, ergs)\,\rm erg~s^{-1}$. 
The required proton acceleration rate  constitutes 1\% of the present spin-down luminosity of the Crab pulsar  
only in the case of the most effective confinement, corresponding to the 
propagation in the extreme Bohm diffusion regime.


\vspace*{1cm}



\newpage

\vfill
\clearpage

\subsection*{Supplementary figures}


\clearpage

\clearpage

\clearpage

\begin{figure}
\centering\includegraphics[width=0.8\linewidth]
{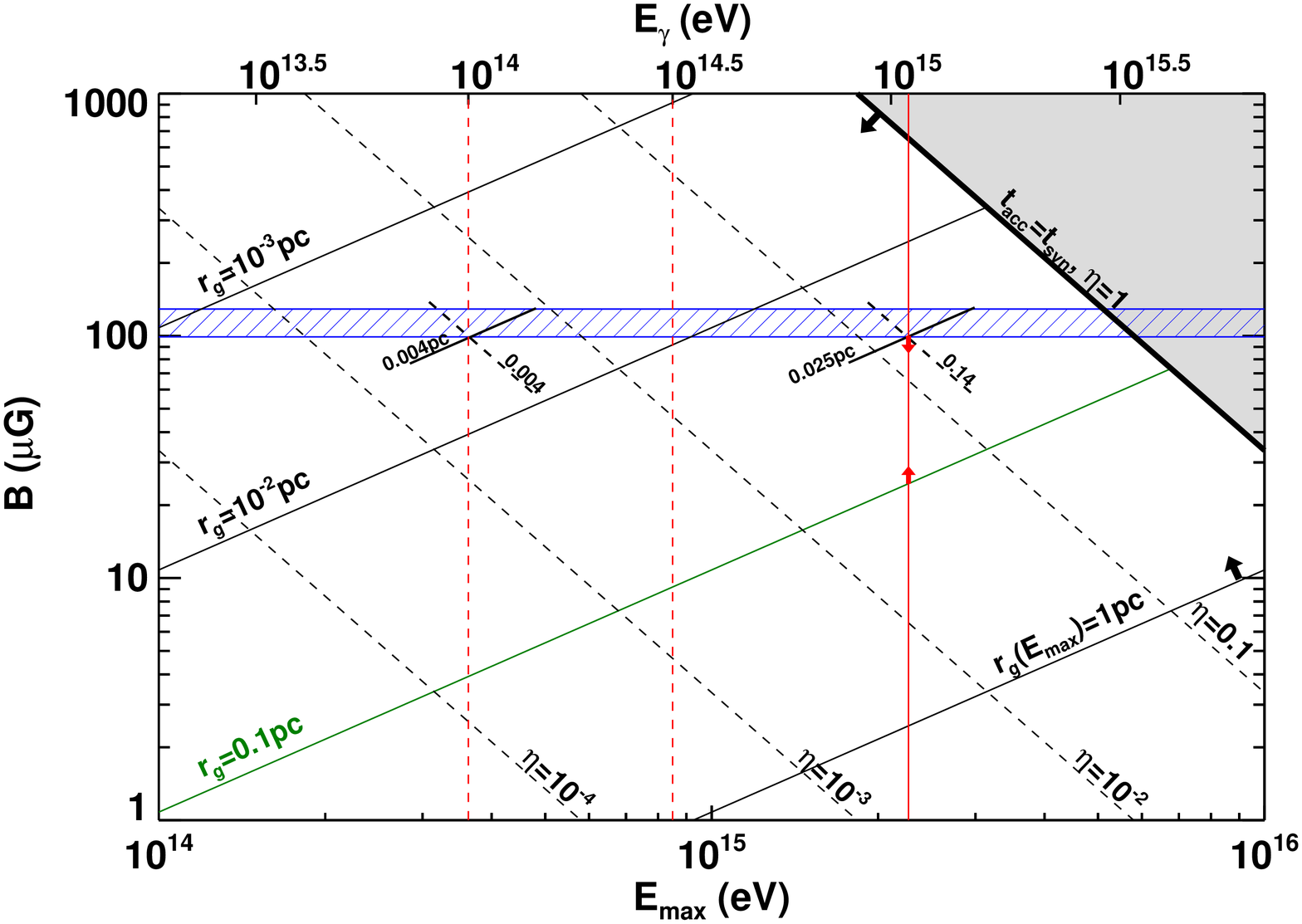}
\caption{{\bf Relation between the magnetic field and the maximum achievable electron energy. }The top horizontal-axis marks the mean energy of the upscattered (from 2.7 K CMBR) IC photon 
energy;  the bottom horizontal-axis represents the energy of parent electrons. The thin solid lines indicate the constraints set by the condition that the larmor radius should not exceed the linear size of the source. The line marked as  0.1~pc,  corresponds to the distance of the position of the electron wind termination  from the pulsar.  The dashed black lines represent the synchrotron cooling-limited maximum electron energies  corresponding to different values of the  acceleration efficiency $\eta$.  The thick solid line corresponds to  $\eta=1$. The shaded 
region is the forbidden parameter space ($\eta >1$). The horizontal blue band corresponds to the magnetic field obtained from the one-zone SED fit shown in Fig.~\ref{SED-onezone}. 
The  solid vertical line 
shows the 1.1\,PeV $\gamma$-ray energy. The short solid and dashed black lines correspond to the lower limits for the size of the acceleration zone and the acceleration efficiency for $B=100 \mu$G at different photon energies.} 
\label{fig:Emax}
\end{figure}

\end{document}